\documentclass[journal]{IEEEtran}
\usepackage{makecell}
\usepackage{tabularx}
\usepackage{algorithm}
\usepackage{algorithmic}
\usepackage{array}
\usepackage{amsfonts}
\usepackage{multirow}
\usepackage{amssymb}
\usepackage{amsmath, amsthm}
\usepackage{mathtools}
\usepackage{bbm}
\usepackage{graphicx}
\usepackage{epstopdf}

\usepackage{epsf}
\usepackage{amsmath, amsfonts}
\usepackage{latexsym}
\usepackage[font=small,labelsep=period]{caption}
\usepackage{subcaption}
\usepackage{multirow}
\usepackage{multicol}
\usepackage{amssymb}
\usepackage{cite}
\usepackage{float}
\usepackage{enumerate}
\usepackage{color}
\usepackage{scrextend}
\usepackage{refcount}
\usepackage{bbm}
\usepackage{lipsum}
\usepackage{mathtools}
\usepackage{cuted}
\usepackage{scalerel}
\usepackage{dsfont}
\newmuskip\pFqmuskip
\usepackage[most]{tcolorbox}
\tcbuselibrary{breakable}
\usepackage{enumitem}
\usepackage{dblfloatfix}
\allowdisplaybreaks

\begin{document}

\title{\fontsize{21}{22}\selectfont Communication and Consensus Co-Design for\\ Distributed, Low-Latency and Reliable Wireless Systems}

\author{
Hyowoon~Seo,~\IEEEmembership{Student~Member,~IEEE},
	Jihong~Park,~\IEEEmembership{Member,~IEEE},
	Mehdi~Bennis,~\IEEEmembership{Senior~Member,~IEEE}
        and~Wan~Choi,~\IEEEmembership{Fellow,~IEEE}
\thanks{Manuscript received January 2, 2020; revised March 24, 2020 and April 27, 2020; accepted May 19, 2020. This research was supported in part by Institute for Information \& communications Technology Promotion(IITP) grant funded by the Korea government(MSIT) (No.2018-0-00809, Development on the disruptive technologies for beyond 5G mobile communications employing new resources), in part by Foundry Division, Device Solution Business, Samsung Electronics Co., LTD, in part by the Academy of Finland project MISSION, in part by the Academy of Finland project SMARTER and in part by the 2019 EU-CHISTERA project LeadingEdge. \emph{(Corresponding author: Wan Choi)}}
\thanks{H. Seo and W. Choi were with the School of Electrical Engineering, KAIST, Korea, and are now with Department of Electrical and Computer Engineering, Seoul National University (SNU), Seoul 08826, Korea (e-mail: hyowoonseo@snu.ac.kr, wanchoi@snu.ac.kr).}
\thanks{J. Park was with the University of Oulu, Finland, and is now with the School of Information Technology, Deakin University, Geelong, VIC 3220, Australia (e-mail: jihong.park@deakin.edu.au).}
\thanks{M. Bennis is with the Centre for Wireless Communications, University of Oulu, Oulu 90014, Finland (email: mehdi.bennis@oulu.fi)}
\thanks{Copyright (c) 2020 IEEE. Personal use of this material is permitted. However, permission to use this material for any other purposes must be obtained from the IEEE by sending a request to pubs-permissions@ieee.org.}}




\maketitle
\vspace{-1cm}
\begin{abstract}
Designing distributed, fast and reliable wireless consensus protocols is instrumental in enabling mission-critical decentralized systems, such as robotic networks in the industrial Internet of Things (IIoT), drone swarms in rescue missions, and so forth. However, chasing both low-latency and reliability of consensus protocols is a challenging task. The problem is aggravated under wireless connectivity that may be slower and less reliable, compared to wired connections. To tackle this issue, we investigate fundamental relationships between consensus latency and reliability through the lens of wireless connectivity, and co-design communication and consensus protocols for low-latency and reliable decentralized systems.  Specifically, we propose a novel communication-efficient distributed consensus protocol, termed \emph{Random Representative Consensus (R2C)}, and show its effectiveness under \emph{gossip} and \emph{broadcast} communication protocols. To this end, we derive a closed-form end-to-end (E2E) latency expression of the R2C that guarantees a target reliability, and compare it with a baseline consensus protocol, referred to as Referendum Consensus (RC). The result show that the R2C is faster compared to the RC and more reliable compared when co-designed with the broadcast protocol compared to that with the gossip protocol.
\end{abstract}


%
\IEEEpeerreviewmaketitle

\newtheorem{mylemma}{Lemma}
\newtheorem{myremark}{Remark}
\newtheorem{mytheorem}{Theorem}
\newtheorem{mydef}{Definition}
\newtheorem{mycor}{Corollary}
\newtheorem{myexample}{Example}
\newtheorem{mydefinition}{Definition}
\newtheorem{myproposition}{Proposition}
\newtheorem{mycases}{Case Study}
\renewcommand*{\themycases}{(\alph{mycases})}
\newcounter{mytempeqncnt}
\newcommand{\tcyan}[1]{{\textcolor{cyan}{#1}}}
\newcommand*\pFq[6][8]{%
  \begingroup 
  \pFqmuskip=#1mu\relax
  \mathchardef\normalcomma=\mathcode`,
  \mathcode`\,=\string"8000
  \begingroup\lccode`\~=`\,
  \lowercase{\endgroup\let~}\pFqcomma
  {}_{#2}F_{#3}{\left[\genfrac..{0pt}{}{#4}{#5};#6\right]}%
  \endgroup
}
\newcommand{\pFqcomma}{{\normalcomma}\mskip\pFqmuskip}

\section{Introduction}
We are currently witnessing an explosive increase of wireless endpoints in smart homes, autonomous vehicles, and modern industrial environments, which warrants a paradigm shift from centralized and rigid architectures towards decentralized and flexible systems~\cite{Schwab2017, Bernard2016}. For example, cyber-physical systems (CPSs) in Industry 4.0 wirelessly interconnect a variety of nodes ranging from mobile devices to sensors and actuators in a decentralized manner while enabling real-time mission-critical control~\cite{Brown2017}, thereby enhancing human and machine safety in manufacturing, inventory tracking, and self-driving vehicles~\cite{Hermann2016}.

Such decentralized systems enable multiple nodes to carry out \emph{valid control actions in a proper order}, taking into account interactions, malfunctions, and adversarial attacks. We address this problem by leveraging principles of distributed ledger technology (DLT), in which every node stores a \emph{ledger} containing a consensual sequence of valid control actions. The candidate actions proposed by multiple nodes are virtually validated at each node via majority rule, by receiving messages of local validation from the other validators. Then, the order of valid actions is determined by a pre-defined consensual ordering policy, based on each action's average validation time.

Designing a \emph{distributed consensus protocol} performing action validation and ordering operation is the prime goal of this paper. Towards supporting mission-critical and real-time tasks over wireless links, the consensus protocol needs to account for wireless system characteristics and thereby optimize its operations under the consensus latency and reliability trade-off. This raises the following two fundamental questions.

\begin{figure}[!t]
\centering
\includegraphics[width= \columnwidth]{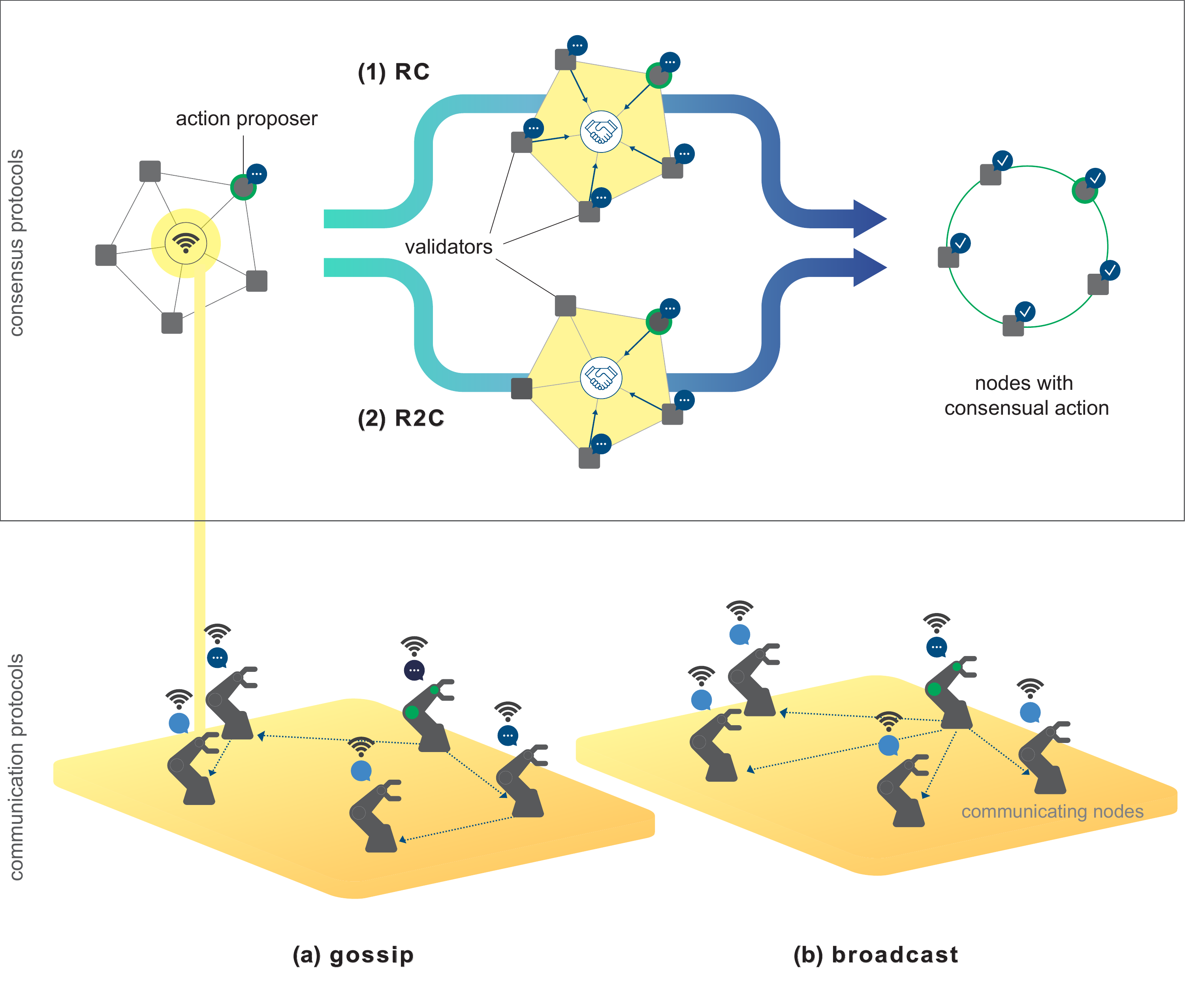}
\caption{\small An illustration of our proposed (2) \emph{random representative consensus (R2C)} protocol compared to (1) a baseline \emph{referendum consensus (RC)} protocol, under (a) \emph{gossip} and (b) \emph{broadcast} communication protocols.}
\label{fig:network}
\end{figure}

\vspace{5pt}\noindent\textbf{Q1}. \emph{How does communication affect the trade-off between consensus latency and reliability?}

For a fixed number of faulty nodes carrying out adversarial and malicious actions, a consensus becomes more reliable when more participating nodes validate actions via consensus. However, too many validators incur huge communication overhead, which increases consensus latency, resulting in the trade-off between consensus latency and reliability.

\vspace{5pt}\noindent\textbf{Q2}. \emph{How to enable distributed, fast and reliable consensus under wireless connectivity?}

To minimize consensus latency, on the one hand, its consensus protocol should be optimized by adjusting the number of validators for guaranteeing a target reliability. On the other hand, the communication protocol of validators should be optimized. In other words, these consensus and communication protocols should be co-designed, under the aforementioned trade-off between consensus latency and reliability.

In this article, we aim to answer \textbf{Q1} and \textbf{Q2} by proposing a novel distributed consensus protocol, termed \emph{Random Representative Consensus (R2C)}. As illustrated in Fig. \ref{fig:network}, in the R2C, only randomly selected representative nodes validate the consensus process. Furthermore, we aim at a communication-efficient design, by investigating the R2C implementations under two different communication protocols: 1) \emph{gossip} based R2C wherein a single message is disseminated through multi-hop communications, and 2) \emph{broadcast} based R2C in which every message is disseminated by a single hop. 

A key design challenge of the R2C is  to obtain the number of representatives ensuring low-latency and high-reliability, which varies depending on the employed communication protocols in general. Thus, we investigate \emph{end-to-end (E2E) latency} and \emph{reliability} of the R2C by deriving their closed-form expressions for the above-mentioned communication protocols. The E2E latency measures the delay that an action experiences between the initiation and completion of its validation. The reliability is studied in terms of \emph{resiliency} against faulty nodes and \emph{robustness} against missing validators.

Compared to a baseline scheme, referred to as \emph{Referendum Consensus (RC)} where all nodes are validators, we show that the R2C can reach its consensus significantly faster, while achieving the target reliability requirement, when reliable direct communication is available between any pair of nodes. More specifically, we analytically find the minimum number of validators required for resilient and robust consensus in the R2C. Moreover, we compare gossip and broadcast communication methods for the RC and R2C, and show that the broadcast-based R2C achieves the fastest consensus, with a sufficiently small amount of faulty nodes. Although broadcasting consumes larger single-hop transmission power than gossiping, the total energy consumption of each consensus is minimized under the broadcast-based R2C, thanks to its lowest consensus latency.

\subsection{Related Work}
The problem of reliability and fault tolerance of consensus protocols has long been studied, mostly under peer-to-peer network architectures with a (relatively) small number of nodes \cite{Lamport1982, Fischer1985, Schneider1990, Castro1999}. However, recent interest in value transfer applications, such as crypto-currencies and smart contracts, has triggered a rapid development of distributed consensus protocols for large-scale systems with low-latency.

In terms of the scalability and reliability, Blockchain is one of the most popular and widely-utilized technology for applications ranging from crypto-currency \cite{Nakamoto} to distributed machine learning \cite{Kim2019} and drone-aided mobile edge computing~\cite{Lee2019}. However, since Blockchain allows permission-less node participation~\cite{Nakamoto}, it may suffer from large consensus delays (e.g., several minutes-hours) that are ill-suited for mission-critical and real-time control applications. 

From the low-latency perspective, permissioned consensus protocols are gradually emerging as suitable alternatives. These methods are built on Byzantine fault tolerant (BFT) algorithms~\cite{Lamport1982, Castro1999} that require exchanging voting information prior to the consensus process, hindering their scalability. In view of this, Hashgraph is one compelling algorithm, in which the consensus is locally carried out at each node without exchanging voting information \cite{Baird2016}, thereby achieving its scalability with low consensus latency.

Nonetheless, most of the aforementioned algorithms postulate that nodes communicate over fast and reliable wired links. To support large-scale systems, wireless connectivity is mandatory in consensus operations, and its impact on consensus reliability and latency should be carefully examined. On this account, wireless distributed consensus protocols have recently been studied in several recent works~\cite{Seo2018,Danzi2019,Liu2019,Al-Jaroodi2019, Lee2019, Kim2019,Nguyen2019,Zhao2019}. For instance, a Hashgraph-motivated wireless distributed consensus protocol has been introduced in \cite{Seo2018}, in the context of distributed wireless spectrum access applications. For power grid applications, an Ethereum-based smart contract and its operation protocol has been studied in \cite{Danzi2018}.

Yet, most of the preceding works on wireless distributed consensus protocols are application-specific, and void of clarifying the relationship between wireless communication and consensus protocol operations. To the best of our knowledge, this work is the first of its kind that investigates the fundamentals of fast and reliable consensus over wirelessly connected nodes via communication and consensus co-design.

\subsection{Contributions and Organization}
The contributions of this paper are summarized as follows.
\begin{itemize}
\item We propose a novel communication-efficient distributed consensus scheme, the R2C (Sec. \ref{sec:R2C}), and compare its effectiveness with a baseline method, RC (Sec. \ref{sec:baseline}). 

\item We derive the minimum required number of the R2C validators under the gossip and broadcast communication protocols, guaranteeing a target resiliency probability against faulty nodes and a target robustness probability against missing validators (\textbf{Propositions}~\textbf{\ref{prop:tolerance}}-\textbf{\ref{thm:reliability}} in Sec. \ref{subsec:resiliency} and C).

\item We derive closed-form E2E latency expressions of the RC and R2C under the gossip and broadcast protocols, guaranteeing a target resiliency and robustness requirement (\textbf{Propositions}~\textbf{\ref{prop:e2egossipRC}}-\textbf{\ref{prop:e2ebroadcastRC}} in Sec. \ref{subsec:baselinelatency} and \textbf{Propositions~\ref{prop:averagegossiplatency}}-\textbf{\ref{prop:e2elatencybroadcastR2C}} in Sec. \ref{subsec:latencyofR2C}).

\item  We provide a distributed consensus and wireless communication co-design guideline, emphasizing that the R2C with wireless broadcast is a faster and more reliable consensus solution for distributed systems when direct communication is available between the nodes composing a network. Its effectiveness and feasibility are underpinned by both analysis and numerical evaluations (Sec. V).
\end{itemize}

The remainder of this paper is organized as follows. In Sec.~II, we explain the system architecture including network model and the communication protocols in detail. In Sec.~III, we study the baseline RC, and in Sec.~IV we propose the R2C. In Sec.~V, we numerically evaluate the effectiveness of the RC and R2C under the gossip and broadcast communication protocols, followed by the conclusion in Sec.~VI.

\section{System Model}\label{sec:systemdescription}
In this section, we describe the network model and communication protocols under study. The communication protocol incorporates two types of message disseminating protocols: gossiping and broadcasting. Some important notations are summarized in Table I for the sake of convenient reference.

\begin{table}[!t]
\renewcommand{\arraystretch}{1}
\captionsetup{justification=centering, labelsep=newline}
\caption{Summary of Notations}
\label{table:notation}
\centering
\begin{tabularx}{\columnwidth}{l l}
\hline
\textbf{Notation}            & \textbf{Meaning} \\
\hline
$N$, $\tilde{N}$ & \# of validators in RC and R2C, respectively.\\
$F$, $\tilde{F}$ & \# of faulty nodes in RC and R2C, respectively.\\
$H_{ik}$ & Channel gain from node $i$ to $k$ $(\sim \mathcal{CN}(0,P_t))$.\\
$R_{ik}$, $R$  & \makecell[l]{Distance between node $i$ and $k$, and any two neighboring \\ nodes, respectively.}\\
$P_{t,g}$, $P_{t,b}$  & \makecell[l]{Transmit power utilized at each node in gossip and \\ broadcast protocols, respectively.}\\
$\tau$ & Time span of a single time slot.\\
$L_g$, $L_b$ & E2E latency in RC with gossip and broadcast, respectively.\\
$\tilde{L}_g$, $\tilde{L}_b$ & E2E latency in R2C with gossip and broadcast, respectively.\\
$w_{i,\zeta}$ & Dissemination time duration of node $i$.\\
$\alpha$ & Target resiliency probability.\\
$\beta$, $\gamma$ & \makecell[l]{Acceptable consensus distortion and target robustness \\ probability, respectively.}\\
$\zeta$ & Target dissemination success probability.\\
\hline
\end{tabularx}
\end{table}

\subsection{Network and Channel Model}\label{subsec:networkchannelmodel}
Considering a network consists of a set $\mathcal{N}$ of $N + 1$ static nodes that are placed in a square grid is sufficient to show the effectiveness of the proposed consensus protocol and facilitates tractable analysis. The network under study  is assumed to be \emph{permissioned} \cite{Lamport1982, Androulaki2018}, in which each node is aware of the identities of the other nodes and size of the network. The nodes can be interpreted as arbitrary network edges ranging from mobile devices to sensors, actuators, and controllers in Industry 4.0. The coordinates of the node $i$, for $i\in\mathcal{N}$, are denoted by $(x_i, y_i)$, and the distance between the nodes $i$ and $k$ is thereby given as $R_{ik} = R_{ki} = \sqrt{(x_i - x_k)^2 + (y_i - y_k)^2}$. For the sake of convenience, we hereafter consider case when $\sqrt{N+1}$ is a positive integer.

The nodes are equipped with wireless transceivers that utilize  equal transmission power $P_t$, and communicate with the other nodes over wireless channels. The wireless channels are assumed to follow the standard path loss model and Rayleigh fading model~\cite{Goldsmith2005}. Specifically, the path loss between nodes $i$ and $k$ is given as
\begin{align}\label{eq:pathloss}
\mathrm{PL}_{\mathrm{dB}}(R_{ik}) = \mathrm{PL}_{\mathrm{dB}}(R_0) + 10 \eta \log_{10} \left(\frac{R_{ik}}{R_0}\right),
\end{align}
where $\mathrm{PL}_{\mathrm{dB}}(R_0)$ is defined as the path loss at the reference distance $R_0$, $\lambda$ is the wavelength and $\eta\geq 2$ indicates the path loss exponent. Furthermore, the multi-path effect on the channel from node $i$ to node $k$ is characterized by the Rayleigh fading model. Let $H_{ik}$ be the channel gain from node $i$ to node $k$, $i \neq k$, following the complex normal distribution with zero mean and variance $P_t$ (i.e., $H_{ik} \sim \mathcal{CN}(0,P_t)$). We assume that the channel gains are independently and identically distributed (i.i.d.). Consequently, when a signal is transmitted from node $i$ to $k$ with transmit power $P_t$, the signal-to-noise ratio (SNR) is represented as
\begin{align}
\mathrm{SNR}_{ik} = 10^{-\frac{\mathrm{PL}_{\mathrm{dB}}(R_0)}{10}} \frac{|H_{ik}|^2 P_t}{P_{n}}\left(\frac{R_0}{R_{ik}}\right)^{\eta},
\end{align}
where $P_{n}$ is the additive white Gaussian noise (AWGN) power.
The absolute time is globally synchronized periodically with GPS \cite{Mahmood2018, Calder2007} and split into time slots of fixed time interval such that
\begin{align}\label{eq:timeslot}
\tau = \frac{M}{B\log(1+\rho)}\ \textrm{seconds},
\end{align} 
where $M$ is the maximum size of message sent by nodes during the consensus protocol in bits, $\rho$ is the target signal-to-noise (SNR) of each transmission, and $B$ in Hz denotes the bandwidth utilized for transmission. An SNR outage occurs if SNR is below $\rho$. Since $|H_{ik}|^2$ is exponentially distributed, the SNR outage probability is given as
\begin{align}\label{eq:snroutage}
\epsilon_{ik} &= 1 - \exp\left( - 10^{\frac{\mathrm{PL}_{\mathrm{dB}}(R_0)}{10}} \rho\frac{P_{n}}{P_t} \left(\frac{R_{ik}}{R_0}\right)^{\eta} \right).
\end{align}
For each outage event, we consider the type-I hybrid automatic repeat request (HARQ), where the message transmissions are repeated until the first success.

\subsection{Communication Protocol}\label{subsec:protocols}
During consensus operations, every node can become either a message source or its destination. Each message is then disseminated from a single \emph{source} to multiple \emph{destinations}, according to either the gossip or broadcast communication protocol, as detailed next.

\subsubsection{Gossip Protocol}\label{subsubsec:gossip}
The nodes make use of multi-hop communication for disseminating messages using the gossip protocol. Unlike a typical gossip protocol in peer-to-peer wired networks, the spread of information is constrained by wireless communication coverage. Thus, we consider a neighbor gossip protocol, where direct communications are only available to the nodes located within the coverage of a transmitter. Specifically, in the network model under study, neighbors are located $R$ meters apart from each other. Assume that all transmit nodes including the source and relays utilizes the same transmit power $P_t = P_{t,g}$, which can be fairly small since we assume communication between neighbors. 

\subsubsection{Broadcast Protocol}
Nodes communicates with other nodes within a single hop in the broadcast protocol, which means that any pair of nodes can be paired through a wireless channel. Compared to the gossip protocol, each transmit node in the broadcast protocol must cover larger areas, and thus, we assume that the transmission power $P_t = P_{t,b}$ used by each node in the broadcast protocol is larger than that of the gossip protocol, i.e., $P_{t,b} \geq P_{t,g}$.

Later on, above two communication protocols are co-designed with the consensus protocols and the consequential performances are compared. Notice that since we are considering a permissioned network with static nodes, the communication protocol that suits better to the given condition, e.g., network size, transmission power, etc., can be chosen right after the formation of the network. For real world implementation with more general network environments, such as network expanding as the number of node increases, the selection of communication protocol can be done after device discovering procedure for device-to-device communication coordination. The decision rule for selecting an appropriate communication protocol can be designed based on the number of other device in its communication range learned from the discovering process.

In the mean time, for the efficient use of radio resources, we preallocate $w_i\tau$ seconds of \emph{dissemination time duration} for message dissemination by node $i \in \mathcal{N}$ as a source. Let $W_i$ be a random variable denoting the number of time slots required for message dissemination to all destinations from source $i$. In both protocols, $w_i$ is determined such that
\begin{align}\label{eq:zeta}
\Pr[W_i \leq w_{i}] \geq \zeta,
\end{align} 
for some target dissemination success probability $0 \leq \zeta < 1$.
%

\subsection{Distributed Ledger Architecture}
Every node is equipped with a sufficiently large storage capacity to store a chain of valid actions, namely a \emph{distributed ledger}. Throughout the paper, we suppose that the distributed ledger is a replicated state machine \cite{Schneider1990} that takes validity of the proposed actions and their timestamps recorded during the validation processes as an input, and outputs a series of valid actions. 
Note that the distributed ledgers are updated and synchronized by the consensus protocol. Moreover, we use cryptographic techniques to prevent fabrication of the messages and detect corrupted messages. We assume that the messages contain public-key signatures \cite{Rivest1978}, message authentication codes \cite{Tsudik1992}, and message digests produced by collision-resistant hash functions \cite{Rivest1992}. Throughout the paper, a message $\mathbf{M}$ signed by node $i$ is denoted by $[\mathbf{M}]_{i}$. Note that a method of signing a digest of a message and appending it to the raw message is widely used rather than signing the full message in practice. Thus, the raw message $\mathbf{M}$ and the encrypted digest of $\mathbf{M}$ are included in the signed message $[\mathbf{M}]_i$.

\section{Baseline: Referendum Consensus (RC)}\label{sec:baseline}
In this section, we introduce our baseline scheme, the referendum consensus (RC) protocol, which is named after a political term \emph{referendum}. The RC aims to reach two kinds of consensual decisions on the validity of proposed actions and the order of valid actions, rooted in the Practical Byzantine Fault Tolerance (PBFT)~\cite{Castro1999} and a permissioned DLT~\cite{Baird2016}, respectively. To this end, all nodes in the RC become validators and participate in the consensus process, as detailed next.

In the RC, each node plays one of the following roles:
\begin{itemize}
\item A \emph{proposer} who proposes a new action to the validators; or
\item A \emph{validator} who validates the proposed actions and shares the validated result with other validators to determine whether to accept the proposed action.
\end{itemize}
For disinterested validation, it is assumed that the proposer does not engage in validating the proposed action of its own and all nodes except for the proposer are validators in the RC. In addition, the system is aware of the number of faulty nodes $F\ (\leq N)$, which can be learned from past consensus rounds. To delve into the fundamentals of co-designing consensus and communication protocols, hereafter we focus on a single proposed action and its consensus process operated over wireless channels with frequency bandwidth $B$. We believe that our key findings are applicable for multiple proposed actions with minor modifications, e.g., by incorporating orthogonal frequency-division multiple access (OFDMA) with the conventional Listen-Before-Talk method or distributed spectrum access (DSA) with the Consensus-Before-Talk algorithm in our prior work \cite{Seo2018}.

\subsection{Operational Structure of RC }

\begin{figure}[!t]
\centering
\begin{subfigure}{0.495\textwidth}
  \centering
  \includegraphics[width=\textwidth]{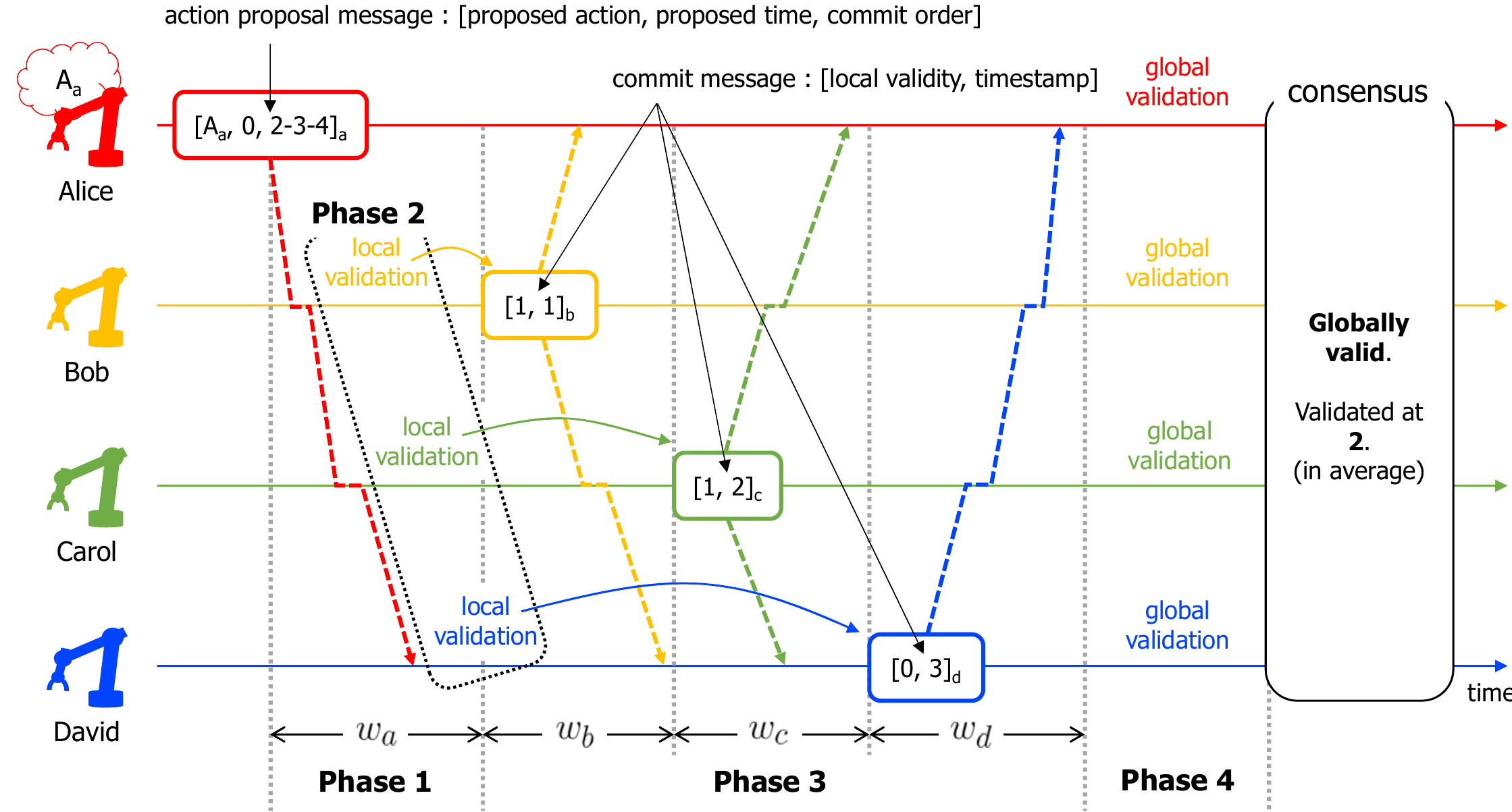}
  \caption{}
  \label{fig:gossipRC}
\end{subfigure}
\begin{subfigure}{0.495\textwidth}
  \centering
  \includegraphics[width=\textwidth]{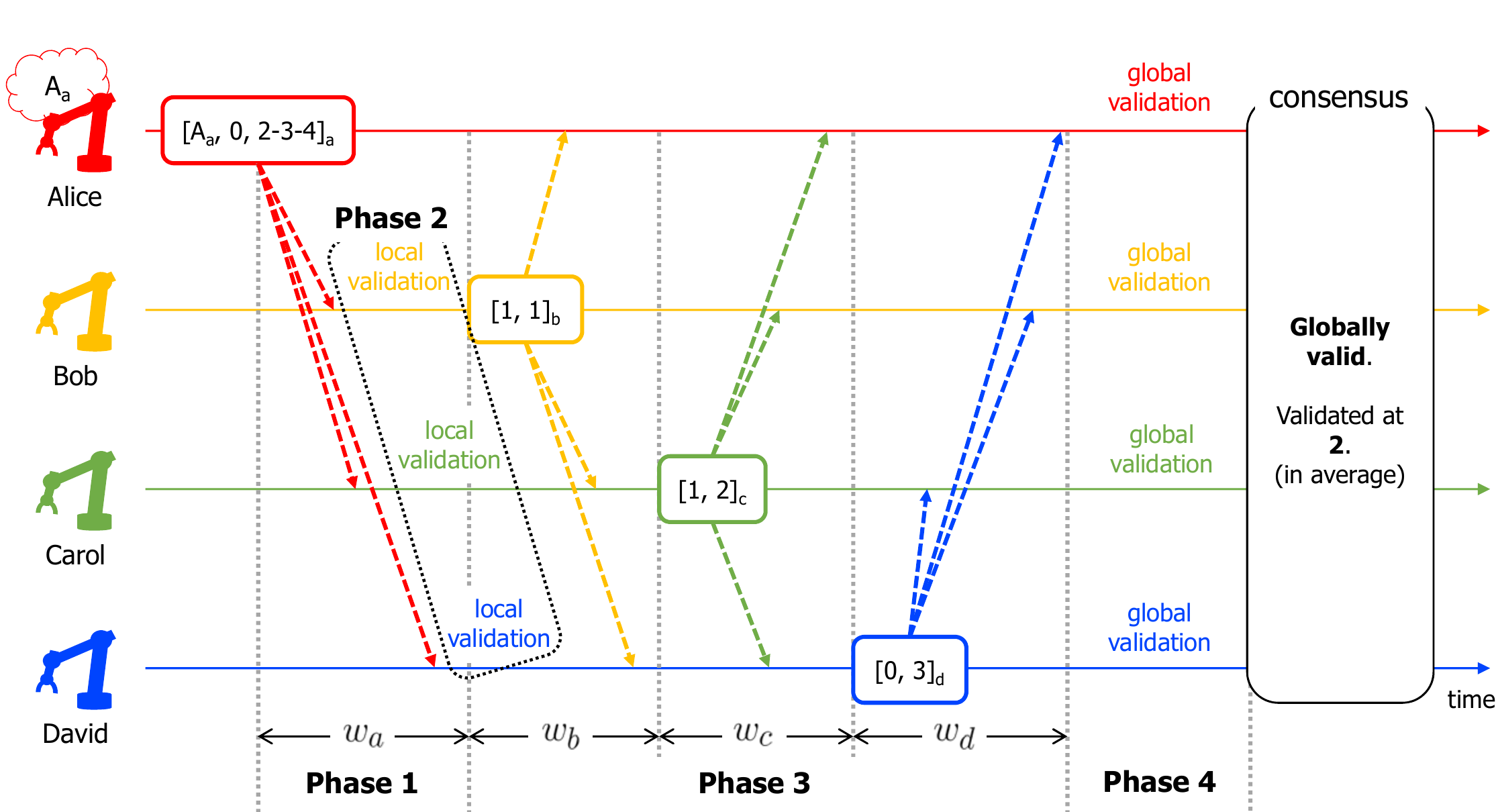}
  \caption{}
  \label{fig:broadcastRC}
\end{subfigure}
\caption{\small  Examples of the RC with (a) gossip and (b) broadcast protocols in the network composed of 4 nodes (Alice, Bob, Carol and David), where Alice is the proposer and the rest are validators.}
\label{fig:RC}
\end{figure}

The RC operates using either the gossip or broadcast communication protocol, as visualized in Fig.~\ref{fig:RC}. For both the cases, there are four operational phases:

\noindent \textbf{Phase 1} \emph{(Action Proposal)} Suppose node $p \in \mathcal{N}$ is a proposer and denote the action proposed by the proposer $p$ as $A_p$. In addition, define the set of validators as $\mathcal{V}_p = \mathcal{N}\backslash\{p\}$, which means that all nodes except for the proposer are validators. In this phase, the proposer becomes a source and the validators become destinations from a communication perspective. The proposer first selects a vacant frequency band of bandwidth $B$ and initiates the RC by disseminating a signed action proposal message 
\begin{align}
[\mathbf{M}_{\mathrm{proposal}}(A_p, \mathcal{V}_p)]_p = [ A_p, T(A_p)_p, S(\mathcal{V}_p) ]_p,
\end{align}
where the message is a tuple of the proposed action $A_p$, the absolute time when the action is proposed $T(A_p)_p$ and the randomly chosen sequence of validator indices $S(\mathcal{V}_p)$, which inform the validators of the committing order of the validated results in Phase 3. Note that the dissemination time duration given to proposer $p$ is $w_{p}$ time slots.

\noindent \textbf{Phase 2} \emph{(Local Validation)} After receiving the proposed action, each validator goes through a local validation of the proposed action based on information stored at the distributed ledger. A validator determines that the proposed action is \emph{locally valid} if the action does not contradict with other valid actions already stored in the ledger, or \emph{locally invalid} if not. The double spend problem in a crypto-currency system can be a good example for the contradiction between the newly proposed action and the existing actions. The local validity of the proposed action $A_p$ determined by validator node $v \in \mathcal{V}_p$ is denoted by $V(A_p)_v$, which is the binary information that takes $1$ if locally valid, and $0$ if locally invalid. The validator also records a timestamp when it finishes the local validation of the proposed action, and the timestamp recorded at node $v$ is denoted by $T(A_p)_v$. Timestamps of the same proposed action may differ in distinct validator nodes depending on the message dissemination method, channel condition and local computing time.

\noindent \textbf{Phase 3} \emph{(Commit)} After the local validation, the validators disseminate commit messages by taking turns with a Round Robin time-division approach. The committing order is informed by the proposer and is specified in $S(\mathcal{V}_p)$ of the action proposal message. Note that in this phase, the committing validator node becomes a source and the rest of the nodes, i.e., the other validators and the proposer, become destinations. When it is node $v$'s turn to commit, the node disseminates a signed commit message 
\begin{align}
[\mathbf{M}_{\mathrm{commit},v}(A_p)]_v = [V(A_p)_v, T(A_p)_v]_v.
\end{align}  
The dissemination time duration of the commit message given to validator $v$ is $w_{v}$ time slots for all $v \in \mathcal{V}_p$. We assume that the dissemination of the commit messages are done over the same frequency band of which is used for the action proposal.

\noindent \textbf{Phase 4} \emph{(Global Validation and Action Ordering)} When each validator collects more than $N - F$ local validations on the proposed action received from distinct validators, it determines the global validity of the proposed action based on the majority rule. Namely, if there are more collected votes on locally valid than the votes on locally invalid, then the validator determines the proposed action to be globally valid and vice versa. On the one hand, if every node is non-faulty, then the collected votes on the local validity will be either all locally valid or all locally invalid. On the other hand, if there are $F$ faulty nodes that can harm the global validity of the proposed action, as long as the condition $N > 3F$ holds, the system is resilient against the faulty nodes \cite{Castro1999}. 

The globally valid actions are the candidates of the actions that will be recorded on the distributed ledgers. Although the majority of validators vote on locally valid for the proposed action, we must focus on the fact that the validated time of the proposed action may differ at distinct validator nodes. This in turn may cause asynchrony on the order of valid actions between the distinct distributed ledgers, if there are multiple actions that are undergoing consensus processes at similar time frame. Accordingly, the validators should also reach consensus on the order of the globally valid actions, based on the collected timestamps. Particularly in the RC, each validator takes the average of the collected timestamps and if all the timestamps are correctly received, the validators will get the \emph{consensual timestamp} of the proposed action, which is denoted by
\begin{align}\label{eq:baselineC}
C(A_p,\mathcal{V}_p) =\frac{1}{|\mathcal{V}_p|} \sum_{v \in \mathcal{V}_p} T(A_p)_v.
\end{align}
The order of the valid actions are organized based on the consensual timestamps. Note that due to the reorganization of the valid actions based on the timestamps, there might be some contradictory actions that might violate the causal relation of the actions. Such troublesome actions are unaccepted and announced to be retried later on or discarded.

\subsection{E2E Latency of RC } \label{subsec:baselinelatency}
As defined earlier, E2E latency is the time interval from the action proposal to the global validation and action ordering. In order to focus on the impact of  wireless communication, we assume that the local computation load is relatively small compared to the local computation capability and thus the local computing time is negligibly small. Then the E2E latency of the RC is obtained as
  \begin{align}\label{eq:baselinelatency}
L_{\mathrm{RC}} =  \tau\sum_{i = 1}^{N+1} w_{i},
\end{align} 
with a communication success probability larger than or equal to $\zeta^{N+1}$, for some $0 \leq \zeta < 1$. Note that \eqref{eq:baselinelatency} comes because a single round of the RC consists of $N+1$ turns of independent message dissemination opportunities. The terms $w_i$ and $\zeta$ come from \eqref{eq:zeta}.

\subsubsection{Gossip-based RC }

In the gossip-based RC, sources disseminate messages to destinations via the gossip protocol described in Sec. \ref{subsec:protocols}. In the considering square network composed of $N+1$, the E2E latency of the gossip-based RC is lower bounded as follows.
\begin{myproposition}\label{prop:e2egossipRC}
The E2E latency of the gossip-based RC can be lower bounded as
\begin{align}\label{eq:Lg}
L_{\mathrm{RC},g} \! \geq \!   \begin{cases} \frac{(3\sqrt{N\!+\!1}\! -\!2)(N\!+\!1) - \sqrt{N\!+\!1}}{2} \tau, &  \mathrm{for\ odd}\ \sqrt{N\!+\!1},\\
\frac{(3\sqrt{N\!+\!1} \!-\!2)(N\!+\!1)}{2} \tau, &  \mathrm{for\ even}\ \sqrt{N\!+\!1}.
\end{cases}
\end{align} 
\end{myproposition}
\begin{IEEEproof}
The proof is provided in Appendix \ref{app:proofofe2egossipRC}.
\end{IEEEproof}
Note that the bound \eqref{eq:Lg} is tight with guaranteeing the target dissemination success probability approximately equals to $\zeta \approx 1$, if the SNR outage probability of communication between the neighbors is sufficiently small.

\begin{figure*}[b]
\hrulefill
\setcounter{equation}{14}
\small\begin{align} 
\label{eq:hypergeometric}
\Pr\left[ \tilde{F} < \frac{\tilde{N}}{3} \right] = 1 - \frac{\binom{ \tilde{N} }{\left\lceil \frac{ \tilde{N} }{3} \right\rceil} \binom{N -  \tilde{N} }{F - \left\lceil \frac{ \tilde{N} }{3} \right\rceil}}{\binom{N}{F}} \pFq{3}{2}{1,\left\lceil \frac{ \tilde{N} }{3} \right\rceil - F, \left\lceil \frac{ \tilde{N} }{3} \right\rceil - \tilde{N} }{\left\lceil \frac{ \tilde{N} }{3} \right\rceil + 1,N + \left\lceil \frac{ \tilde{N} }{3} \right\rceil  +1 - F - \tilde{N} }{1},
\end{align}
\setcounter{equation}{10}
\normalsize
\end{figure*}

\subsubsection{Broadcast-based RC}
In the broadcast-based RC, sources disseminate messages to destinations via the broadcast protocol described in Sec. \ref{subsec:protocols}. In the considered network model, we have the following E2E latency of the broadcast-based RC.
\begin{myproposition}\label{prop:e2ebroadcastRC}
The E2E latency of the broadcast-based RC is
\begin{align}
L_{\mathrm{RC},b} = \sum_{i = 1}^{N+1} \left\lceil \frac{\log\left( 1 \! - \!\zeta^{\frac{1}{N}}\right)}{\log{\epsilon_{i,\max}}} \right\rceil \tau,
\end{align} 
with an overall communication success probability larger than $\zeta^{N+1}$, for $0 \leq \zeta<1$, where $\epsilon_{i,\max}$ denotes the maximum among all SNR outage probabilities between node $i$ and all the other nodes.
\end{myproposition}
\begin{IEEEproof}
The proof is provided in Appendix \ref{app:proofe2ebroadcastRC}.
\end{IEEEproof}
Similar to the gossip protocol, for sufficiently small $\epsilon_{i,\max}$, the dissemination outage probability will approach to zero. Note that the  lower bounds of E2E latency for the RC obtained in Propositions 1 and 2 will be compared with the closed-form expression of the E2E latency of the proposed R2C in the next section.


\section{Proposed: Random Representative Consensus~(R2C)}\label{sec:R2C}
In this section, we propose the random representative consensus (R2C) protocol which reduces the E2E consensus latency while guaranteeing a target reliability. As discussed in \textbf{Q1}, too many validators incur long consensus latency, whereas too small validators hinder the consensus reliability. Balancing between latency and reliability, the R2C seeks the minimum number of validators to achieve a target reliability, thereby reducing the consensus latency, as elaborated in the following subsections.

\subsection{Operational Structure of R2C}

For a given proposer node $p$, we assume $\tilde{N}$ \emph{representative} nodes out of $N$ nodes act as a validator, while the other $N-\tilde{N}$ nodes act as \emph{acceptors}, who do not validate the proposed actions, but only aggregate the validated results and determine whether to accept or reject the proposed action.

The R2C operations follow the same procedures of the RC in Sec.~\ref{sec:baseline}, except for the following changes at each phase.

\noindent \textbf{Phase 1} \emph{(Action Proposal)} At first, the proposer $p$ uniformly and randomly selects $\tilde{N}$ representative validators from the set $\mathcal{V}_p$. We define the chosen representative subset as $\tilde{\mathcal{V}}_p$. Then the proposer initiates the consensus protocol by disseminating the action proposal message $[\mathbf{M}_{\mathrm{proposal}}( A_p, \tilde{\mathcal{V}}_p)]_p = [A_p, T(A_p)_p, S(\tilde{\mathcal{V}}_p)]_p$ to the network.

\noindent \textbf{Phase 2} \emph{(Local Validation)} Unlike the RC, the local validation is only done by the members of $\tilde{\mathcal{V}}_p$ in the R2C.

\noindent \textbf{Phase 3} \emph{(Commit)} The members in $\tilde{\mathcal{V}}_p$ take turn to commit the validated results based on the commit order $S(\tilde{\mathcal{V}}_p)$ informed by the proposer. The results are delivered to all members including the proposer, validators and acceptors.

\noindent \textbf{Phase 4} \emph{(Global Validation and Action Ordering)} All members including the proposer, validators, and acceptors go through a global validation process, as in the RC based on the majority rule. The global validation is based on the locally validated results delivered from the representative validators. The consensual timestamp of the R2C is
\begin{align}\label{eq:representativeC}
C(A_p, \tilde{\mathcal{V}}_p) = \frac{1}{|\tilde{\mathcal{V}}_p|}\sum_{v \in \tilde{\mathcal{V}}_p} T(A_p)_v.
\end{align} 

Due to its missing set of validators compared to the RC, the reliability of the R2C should be more carefully examined. For this reason, we study the resilience of the R2C against faulty nodes and its robustness against missing validators in the following subsections.

\subsection{Resiliency of R2C}\label{subsec:resiliency}
Against $F$ faulty nodes, the baseline RC becomes resilient if the number $N$ of validators satisfies $N > 3F$~\cite{Castro1999}. Likewise, the R2C becomes resilient if the number $\tilde{N}$ of representative validators satisfies $\tilde{N} > 3\tilde{F}$. Due to the randomly selected representatives, the resilience of the R2C is guaranteed stochastically. For the resilience outage probability $\alpha$, the  definition of resilience for the R2C is described as below.
\begin{mydef}\label{def:faulttolerant}
For a fixed number of representatives $\tilde{N}$ and  a random number of faulty representative nodes $\tilde{F}$, the R2C is $\alpha$-resilient if 
  \begin{align}
\Pr[\tilde{N} > 3\tilde{F}] \geq \alpha,
\end{align} 
for a target resiliency probability  $\alpha$ where $0 < \alpha \leq 1$.
\end{mydef}

Next, we characterize the minimum number of representative validators for achieving $\alpha$-resiliency. Since the representatives are chosen uniformly by the proposer, $\tilde{N}$ is a random variable which follows the hypergeometric distribution, of which  probability mass function is given by
\begin{align}
\Pr [ \tilde{F} = f ] = \frac{\binom{F}{f} \binom{N-F}{\tilde{N}-f}}{\binom{N}{\tilde{N}}}, \forall f \in \{0,\dots,F\}.
\end{align} 
Accordingly, the resiliency outage probability can be equivalently expressed as \eqref{eq:hypergeometric}, where $_3F_2[\cdot]$ is the generalized hypergeometric function.

\begin{figure*}[b]
\hrulefill
\setcounter{equation}{17}
\small\begin{align}\label{eq:erfinverse}
g^{-1}(x) &= \left[ -\frac{2}{\pi a} - \frac{\log(1-x^2)}{2} + \sqrt{\left( \frac{2}{\pi a} + \frac{\log(1-x^2)}{2} \right)^2 - \frac{1}{a} \log(1-x^2)} \right]^{\frac{1}{2}},\ 0 \leq x <1.
\end{align}
\setcounter{equation}{15}
\normalsize
\end{figure*}

\begin{figure}[!t]
\centering
\includegraphics[width= \columnwidth]{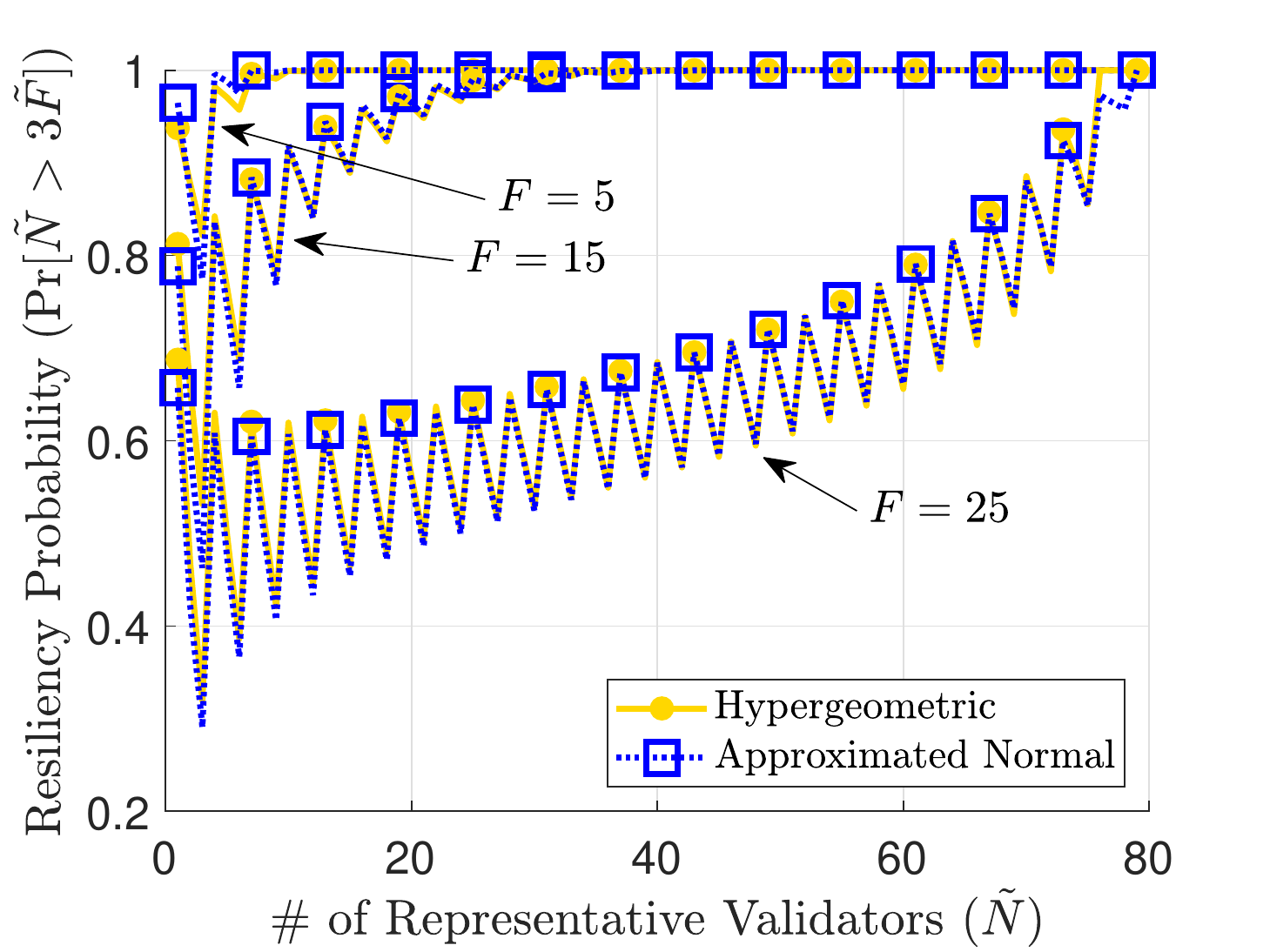}
\caption{\small Illustration of resiliency probability versus the number of representative validators obtained from hypergeometric and approximated normal approach, when $N = 80$ and $F = 5, 15, 25$.}
\label{fig:gossip}
\end{figure}

A straightforward way of obtaining the condition on $\tilde{N}$ for $\alpha$-resiliency is to compute the inverse function of \eqref{eq:hypergeometric}, however, since it includes a hypergeometric function, it is less tractable. Alternatively, we can take advantage of the fact that the hypergeometric distribution can be approximated to the normal distribution when $\tilde{N}$ is sufficiently large, $N$ and $F$ are large compared to $\tilde{N}$, and $\frac{F}{N}$ is not close to $0$ or $1$. Thus, from the normal approximation, \eqref{eq:hypergeometric} yields  
\begin{align}\label{eq:hypertonormal}
\Pr \left[\tilde{F} < \frac{\tilde{N}}{3} \right] \approx \frac{1}{2} \left[ 1 + \mathrm{erf}\left( \frac{\frac{\tilde{N}}{3} - \mu_{\tilde{F}} - \phi}{\sigma_{\tilde{F}}\sqrt{2}} \right) \right],
\end{align} 
where $\mathrm{erf}(x) = \frac{2}{\sqrt{\pi}} \int_{0}^{x} e^{-t^2} \mathrm{d}t$ is the error function, $\mu_{\tilde{F}} = \frac{F\tilde{N}}{N}$ and $\sigma_{\tilde{F}} = \sqrt{ \frac{F\tilde{N}}{N} \frac{N-F}{N}\frac{N-\tilde{N}}{N-1} }$ are the mean and standard deviation of the hypergeometric random variable $\tilde{F}$, respectively, and $\phi$ ($0 < \phi < 1$) is the correction factor that comes from the approximation of the probability function of a discrete random variable to a continuous random variable. In \cite{Winitzki2003,Winitzki2008}, an approximation for the error function $\mathrm{erf}(x)$, whose derivation is based on a generalization of Hermite-Pade approximation, is given as
  \begin{align}\label{eq:approximatederrorfunction}
\mathrm{erf}(x) \approx g(x) = \left[ 1 - e^{-x^2 \frac{\frac{4}{\pi} + ax^2}{1+ax^2}} \right]^{\frac{1}{2}},\ x \geq 0,
\end{align}
where the constant $a \approx 0.14$ is chosen to achieve a relative precision better than $0.004$ uniformly for all real $x \geq 0$. For $x <0$, the identity $\mathrm{erf}(-x) = -\mathrm{erf}(x)$ can be used. Note that the approximated error function $g(x)$ in \eqref{eq:approximatederrorfunction} can be easily inverted analytically as \eqref{eq:erfinverse}. Therefore, we approximate the inverse of the error function as $\mathrm{erf}^{-1}(x) \approx g^{-1}(x)$. Note that for the range of $-1 < x \leq 0$, the identity $\mathrm{erf}^{-1}(x) = -\mathrm{erf}^{-1}(x)$ can also be used.  In Fig. \ref{fig:gossip}, we compare the resiliency probabilities obtained from two different approaches, i.e., the original hypergeometric distribution and approximant with approximated error function. As shown in the figure, the approximation is tight for different number of faulty nodes $F$ and the number of random representative validators $\tilde{N}$. Resultingly, $\tilde{N}$ that approximately achieves $\alpha$-resiliency of the R2C as follows.

\setcounter{equation}{18}
\begin{myproposition}\label{prop:tolerance}
$\alpha$-resiliency of the R2C can be approximately achieved if the number of representatives $ \tilde{N}$ satisfies the following condition
\begin{align}\label{eq:tolerancebound}
\tilde{N} > N_{\alpha},
\end{align} 
where $N_{\alpha} =  \frac{\phi A + BN + \sqrt{2\phi ABN - 2 \phi^2 B + B^2 N^2}}{A^2+2B}$, $A = \frac{1}{3} - \frac{F}{N}$ and $B = \frac{F(N-F)}{(N-1)N^2}(g^{-1}(2\alpha-1))^2$.
\end{myproposition}
\begin{IEEEproof}
From \eqref{eq:hypertonormal} and \eqref{eq:approximatederrorfunction}, we have the condition 
  \begin{align}
\frac{1}{2} \left( 1 + g\left( \frac{\tilde{N}/{3} - \mu_{\tilde{F}} - \phi}{\sigma_{\tilde{F}}\sqrt{2}} \right) \right) \geq \alpha,
\end{align}
achieving $\alpha$-resiliency. By rearranging this to satisfy the condition for $\tilde{N}$, we have \eqref{eq:tolerancebound}.
\end{IEEEproof}
Note that $N_{\alpha}$ depends on $N$, $F$, and $\alpha$. Thus, the number of random representative validators can be easily chosen to achieve $\alpha$-resilience if we have the knowledge of $N$ and $F$.

\subsection{Robustness}\label{subsec:reliability}
We also seek for the condition that ensures the robustness of the R2C against missing validators, especially by measuring the gap between the consensual timestamps of the valid actions in the R2C and that of the RC. We first define a consensus distortion function \cite{Cover2012} as
\begin{align}\label{eq:distortionfunction}
|D(A_p, \mathcal{V}_p, \tilde{\mathcal{V}}_{p})| = | C(A_p, \mathcal{V}_p) - C(A_p, \tilde{\mathcal{V}}_{p})|,
\end{align}
which measures the absolute difference between the RC consensual timestamp $C(A_p, \mathcal{V}_p)$ in \eqref{eq:baselineC} and the R2C consensual timestamp $C(A_i, \tilde{\mathcal{V}}_p)$ in \eqref{eq:representativeC}.
Both $C(A_p,\mathcal{V}_p)$ and $C(A_p, \tilde{\mathcal{V}}_{p})$ are random values, where the randomness of $C(A_p,\mathcal{V}_p)$ is due to the channel uncertainty and that of $C(A_p, \tilde{\mathcal{V}}_{p})$ comes from the randomly chosen representative validator set as well as the channel uncertainty. Intuitively, if the distortion is small, it means that the representatives are well representing the consensual timestamp of the entire network, so that the valid action ordering by the random representative validators will be the same as the ordering done by the all the participants in the network. On the other hand, if the distortion is large, the ordering of the valid actions by the representatives might be different to that done by all nodes. In this context, we define $(\beta,\gamma)$-robustness of the R2C as follows.
\begin{mydef}
The R2C is $(\beta, \gamma)$-robust if
  \begin{align}\label{eq:deltaepsilon}
\Pr[|D(A_p, \mathcal{V}_p, \tilde{\mathcal{V}}_{p})| \leq \beta ] \geq \gamma,
\end{align} 
for an acceptable consensus distortion $\beta$, where $\beta \geq 0$, and target robustness probability $\gamma$, where $0 \leq \gamma \leq 1$.
\end{mydef}
\noindent Note that the number of validators should be sufficiently large, in order to ensure the distortion smaller than $\beta$. Again, we seek for the condition of $\tilde{N}$ that guarantees $(\beta,\gamma)$-robustness of the R2C. A straightforward way of obtaining the condition is to derive the exact distribution of $D(A_p,\mathcal{V}_p, \tilde{\mathcal{V}}_p)$. However, as mentioned before, $D(A_p, \mathcal{V}_p, \tilde{\mathcal{V}}_p)$ is a jointly distributed random variable, where the randomness comes from the random selection of the representative set $\tilde{\mathcal{V}}_p$ and the number of transmissions required for the successful information delivery, so the distribution is complicated and hard to express in a tractable form.

\begin{figure}[!t]
\centering
\includegraphics[width= \columnwidth]{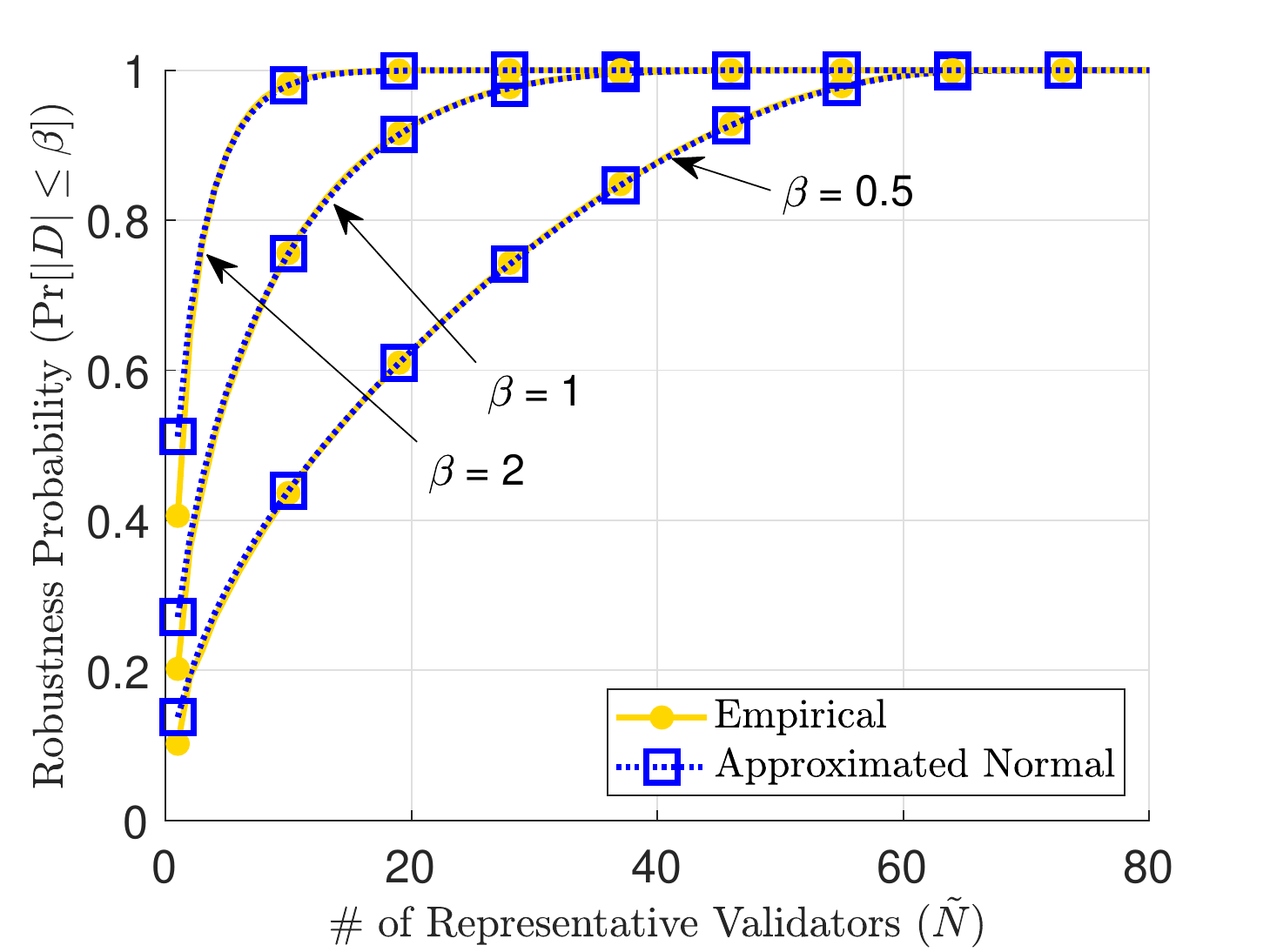}
\caption{\small  Bound on the consensus failure probability due to time stamp distortion vs. the number of random representative validators $\tilde{N}$, obtained from empirical experiment, normal and approximated normal approaches, when $N = 80$ and $\beta = 0.5, 1, 2$.}
\label{fig:distortionoutage}
\end{figure}

Alternatively, we approximate $D(A_p, \mathcal{V}_p, \tilde{\mathcal{V}}_p)$ as a normal distribution with zero mean and variance $\sigma_D^2$. A justification of the normal approximation is as follows. First, assume that all nodes determine that the proposed action is valid, that is $V(A_p)_v = 1$ for all $v \in \mathcal{V}_p$. Then, we can write
  \begin{align}\label{eq:distortionandtime}
D(A_p, \mathcal{V}_p, \tilde{\mathcal{V}}_{p}) = \frac{1}{N} \sum_{i \in \mathcal{V}_p} T(A_p)_i - \frac{1}{\tilde{N}_p} \sum_{j \in \tilde{\mathcal{V}}_p} T(A_p)_j.
\end{align}  
Consequently, for a given some realization $T(A_p)_v = t_v$ for all $v \in \mathcal{V}_i$, the consensual timestamp of the RC is defined as $C(A_p,\mathcal{V}_p) \mid \mathcal{T} $ and given by $C(A_p,\mathcal{V}_p) \mid \mathcal{T} = \frac{1}{N} \sum_{v \in \mathcal{V}_p} t_v$ which can be seen as a population mean over a set $\mathcal{T} = \{t_v \mid v \in \mathcal{V}_p \}$. On the other hand, the consensual timestamp of the R2C $C(A_p,\tilde{\mathcal{V}}_p) \! \mid \! \mathcal{T} = \frac{1}{\tilde{N}_p} \sum_{v \in \tilde{\mathcal{V}}_p} t_v$ is a sample mean where the samples are chosen over the set $\mathcal{T}$ without replacement. It is known that from the Central Limit Theorem (CLT), $C(A_p,\mathcal{V}_p) \! \mid \! \mathcal{T} - C(A_p,\tilde{\mathcal{V}}_p) \! \mid \! \mathcal{T}$ follows normal distribution if the population of the $\mathcal{T}$ is infinite. Although we assume finite $N$, Fig. \ref{fig:distortionoutage} shows that the approximation is quite tight so long as $N$ is sufficiently large. Thus, we have
  \begin{align}
\Pr[|D(A_i, \mathcal{V}_i, \tilde{\mathcal{V}}_{i})| \leq  \beta ] \approx \mathrm{erf}\left( \frac{\beta}{\sigma_{D}\sqrt{2}} \right).
\end{align} 
Approximating the inverse error function via \eqref{eq:erfinverse}, we provide the following proposition.

\begin{myproposition}\label{thm:reliability}
For a given location of the proposer and the message dissemination method, the variance of $D(A_p, \mathcal{V}_p, \tilde{\mathcal{V}}_p)$ is
  \begin{align}\label{eq:D}
\sigma_{D}^2 = \frac{\tau^2(N-\tilde{N})}{\tilde{N} N^2} \psi,
\end{align} 
and the necessary number of representatives for $(\beta, \gamma)$-robustness is
  \begin{align}\label{eq:reliabilitybound}
\tilde{N} > N_{(\beta,\gamma)},
\end{align} 
where $\psi = \sum_{v\in \mathcal{V}_p} \left( \mathrm{E}[Z_{pv}^2] + \frac{1}{N-1} \sum_{j \in \mathcal{V}_p, j \neq v} \mathrm{E}[Z_{pv}] \mathrm{E}[Z_{pj}]\right)$, $N_{(\beta,\gamma)} = \left[ \frac{1}{N} + \frac{\beta^2 N}{2\tau^2 (g^{-1}(\gamma))^2 \psi}\right]^{-1}$, and $Z_{pv}$ is the number of transmissions required for successful message delivery from node $p$ to $v$.
\end{myproposition}
\begin{IEEEproof}
The proof is provided in Appendix \ref{app:proofofbetagamma}.
\end{IEEEproof}

Unlike the condition for the $\alpha$-resiliency, $N_{(\beta,\gamma)}$ is dependent not only on $N$, $F$, $\beta$ and $\gamma$, but also on the the location of the proposer $p$. Also note that the message disseminating method affects the term $\psi$, thus we write $N_{(\beta,\gamma),g}$ and $N_{(\beta,\gamma),b}$ for the gossip and broadcast protocols in the R2C, respectively.


\subsection{E2E Latency of $\alpha$-Resilient and $(\beta,\gamma)$-Robust R2C}\label{subsec:latencyofR2C}
From Propositions \ref{prop:tolerance} and \ref{thm:reliability}, we approximately achieve $\alpha$-resiliency and $(\beta,\gamma)$-robustness by the R2C when the number of random representative validators satisfy
  \begin{align}
\tilde{N} \geq \max \left( N_{\alpha}, N_{(\beta,\gamma)} \right),
\end{align} 
for the given proposer $p$. 
Meanwhile, since the validators are chosen randomly by the proposer, the consensus latency will vary depending on the chosen validator set. Thus, for the given proposer $p$, the expected consensus latency of the R2C is obtained as
  \begin{align}\label{eq:R2Clatency}
L_{\mathrm{R2C}}= \left[w_{p,\zeta}  +  \frac{\tilde{N}}{N} \sum_{v \in \mathcal{V}_p} w_{v,\zeta}\right]\tau,
\end{align} 
with communication success probability larger than $\zeta^{\tilde{N}+1}$, where $w_p$ and $w_{v}$ are the dissemination time duration as discussed in Sec. \ref{subsec:protocols}, and $\frac{\tilde{N}}{N}$ is from the marginalization taken over all possible representatives subset $\tilde{\mathcal{V}}_p \subset \mathcal{V}_p$ and $|\tilde{\mathcal{V}}_p| = \tilde{N}$.

\subsubsection{Gossip-based R2C}
Now we find the number of validators $\tilde{N}$ that achieves $\alpha$-resiliency and $(\beta,\gamma)$-robustness in the gossip-based R2C to obtain E2E latency for given proposer $p$. Suppose the proposer $p$ proposes an action $A_p$ at time $T(A_p)_p$ and since we are assuming a negligible local computation time at each node, the timestamp of the proposed action at a validator node $v \in \mathcal{V}_p$ can be expressed as
  \begin{align}
T(A_p)_{g,v} &= T(A_p)_p + \tau Z_{g,pv},
\end{align} 
where $Z_{g,pv}$ is the random variable that denotes the number of time slots required for delivery of a message from node $p$ to node $v$. From \eqref{eq:baselineC}, the consensual timestamp of $A_p$ in the gossip-based RC can be expressed as
  \begin{align}\label{eq:C_gossip}
C(A_p,\mathcal{V}_p)_{g} &= T(A_p)_p + \frac{\tau}{N} \sum_{v \in \mathcal{V}_p} Z_{g,pv}\\
&\approx T(A_p)_p + \frac{\tau}{N} \sum_{v \in \mathcal{V}_p} e_{pv},
\end{align} 
and the consensual timestamp in the gossip-based R2C can be expressed as
  \begin{align}
C(A_p,\tilde{\mathcal{V}}_p)_{g} \approx T(A_p)_p + \frac{\tau}{\tilde{N}} \sum_{v \in \tilde{\mathcal{V}}_p}e_{pv},
\end{align} 
where $\tilde{\mathcal{V}}_p$ is the random representative validators set and $e_{pv}$ is the number of edges of the shortest paths from node $p$ to $v$ as defined in Appendix \ref{app:proofofe2egossipRC}.

\begin{figure*}[b]
\hrulefill
\setcounter{equation}{35}
\small\begin{align}\label{eq:tildeLga}
L_{\mathrm{R2C},g}^{\mathrm{cor}} \geq \begin{cases} \left[ \left( \frac{3\sqrt{N+1}}{2} - \frac{\sqrt{N+1}-1}{N} -1 \right) \tilde{N} + 2(\sqrt{N+1} -1)\right]\tau, & \mathrm{for\ odd}\ \sqrt{N+1},\\
\left[ \left( \frac{3\sqrt{N+1}}{2} - \frac{\sqrt{N+1}-2}{2N} -1  \right) \tilde{N} + 2(\sqrt{N+1} -1)\right]\tau, & \mathrm{for\ even}\ \sqrt{N+1},
\end{cases}
\end{align}
\normalsize
\setcounter{equation}{32}
\end{figure*}

From Proposition \ref{thm:reliability}, we can show that the number of representatives in the gossip-based R2C must be no smaller than
  \begin{align}
N_{(\beta,\gamma),g} = \left[ \frac{1}{N} + \frac{\beta N}{2\tau^2 (g^{-1}(\gamma))^2 \psi_{g}} \right]^{-1},
\end{align} 
where $\psi_{g}$ varies depending on the location of the proposer node $p$ to achieve $(\beta,\gamma)$-robustness.

For instance, if the proposer is located at the corner point of the network, we have 
\begin{align}\label{eq:psigp}
\psi_{g}^{\mathrm{cor}} \! = \! \frac{(N\!+\!1)((13N \! - \! 24\sqrt{N \! + \! 1} + 16)N \! + \! 12(\sqrt{N\!+\!1} \! - \! 1))}{6(N-1)},
\end{align}
and if the proposer is located at the center point of the network, we have
\begin{align}
\psi_{g}^{\mathrm{cen}} = \frac{(13N^2 - 4N - 8)N}{24(N-1)}.
\end{align} 
The number of representatives for $\alpha$-resiliency and $(\beta,\gamma)$-robustness in the gossip-based R2C must satisfy $\tilde{N} \geq \max (N_{\alpha}, N_{(\beta,\gamma),g})$ where $N_{\alpha}$ is fixed number for given $N$, $F$ and $\alpha$ as mentioned in Sec. \ref{subsec:resiliency}. We can also derive the E2E latency of the gossip-based R2C as follows.
\begin{myproposition}\label{prop:averagegossiplatency}
The E2E latency of the gossip-based R2C is lower bounded as \eqref{eq:tildeLga}, if the proposer is located at the corner point of the network and 
\setcounter{equation}{36}
\begin{align}\label{eq:tildeLgb}
L_{\mathrm{R2C},g}^{\mathrm{cen}} \geq \left[ \left( \frac{3}{2}\sqrt{N+1} -1 \right)\tilde{N} + \sqrt{N+1}-1\right] \tau,
\end{align} 
if the proposer is located at the center point of the network.
\end{myproposition}
\begin{IEEEproof}
The results follow from the proof of Proposition \ref{prop:e2egossipRC} in Appendix \ref{app:proofofe2egossipRC} with \eqref{eq:R2Clatency}.
\end{IEEEproof}
Similar to the bound \eqref{eq:Lg}, the bounds \eqref{eq:tildeLga} and \eqref{eq:tildeLgb} are tight if the SNR outage probability of a communication between two neighbors is sufficiently small.

\subsubsection{Broadcast-based R2C}
Suppose the proposer $p$ starts disseminating the action proposal message at time $T(A_p)_p$. Then the timestamp of the proposed action at validator $v \in \tilde{\mathcal{V}}_p$ is
  \begin{align}
T(A_p)_{b,v} = T(A_p)_p + \tau Z_{b,pv}.
\end{align} 
Assuming that $V(A_p)_v = 1$, $\forall v \in \mathcal{V}_p$, the consensual timestamp of the proposed action $A_p$ in the RC with the broadcasting is
  \begin{align}\label{eq:C_broadcast}
C(A_p,\mathcal{V}_p)_{b} &= T(A_p)_p + \frac{\tau}{N} \sum_{v \in \mathcal{V}_p}  Z_{b,pv}.
\end{align} 
Similarly, the consensual timestamp in the R2C with the broadcasting  can be expressed as
  \begin{align}
C(A_p,\tilde{\mathcal{V}}_p)_{b} &= T(A_p)_p + \frac{\tau}{\tilde{N}} \sum_{v \in \mathcal{V}_p}  Z_{b,pv}.
\end{align} 
For $(\beta,\gamma)$-robustness, the number of representatives in the broadcast-based R2C must be no smaller than
  \begin{align}
N_{(\beta,\gamma),b} = \left[ \frac{1}{N} + \frac{\beta^2 N}{2 \tau^2 (g^{-1}(\gamma))^2 \psi_{b}} \right]^{-1},
\end{align} 
where $\psi_{b} \! = \! \sum_{v \in \mathcal{V}_p} \! \left( \frac{ 1 + \epsilon_{pv} }{(1\! - \! \epsilon_{pv})^2} \! + \! \frac{1}{N-1}\sum_{j\in\mathcal{V}_p,j\neq v } \frac{1}{(1\!-\!\epsilon_{pv})(1\!-\!\epsilon_{pj})}\right)$.

Thus, the number of representatives for the $\alpha$-resiliency and $(\beta,\gamma)$-robustness in the broadcast-based R2C must satisfy $\tilde{N} \geq \max (N_{\alpha}, N_{(\beta,\gamma),b})$. Moreover, we can derive the E2E latency of the broadcast-based R2C as follows. 
\begin{myproposition}\label{prop:e2elatencybroadcastR2C}
The E2E latency of the broadcast-based R2C is
  \begin{align}\label{eq:tildeLb}
L_{\mathrm{R2C},b} \! = \! \left[ \frac{\tilde{N}}{N}\sum_{i \in \mathcal{V}_p} \left\lceil \frac{\log\left( 1 \! - \! \zeta^{\frac{1}{N}}\right)}{\log{\epsilon_{i,\max}}} \right\rceil \! + \! \left\lceil \frac{\log\left( 1 \! - \! \zeta^{\frac{1}{N}}\right)}{\log{\epsilon_{p,\max}}} \right\rceil \right] \tau,
\end{align} 
with  a communication success probability greater than or equal to $\zeta^{\tilde{N} + 1}$.
\end{myproposition}
\begin{IEEEproof}
The results follow from the proof of Proposition \ref{prop:e2ebroadcastRC} in Appendix \ref{app:proofe2ebroadcastRC} with \eqref{eq:baselinelatency}.
\end{IEEEproof}

\section{Simulation Results}
In this section, we numerically evaluate the performance of the RC and the R2C, and validate the analytic results obtained in the previous sections. Taking into account of the communications between IoT devices, e.g., devices equipped with Bluetooth-based transceivers, we fix the transmit power  at each node for the gossip and the broadcast transmissions as $P_{g,t} = 2.5$ mW and $P_{b,t} = 100$ mW, respectively, and the noise power as $P_{\mathrm{noise}} = 10^{-10}$ mW. For fair comparison, we simulate and compare the total energy consumption for accepting a single proposed action in the consensus protocols under study. In addition, we set the distance between the two neighboring nodes to be $R = 10$ meters and the target dissemination success probability to be $\zeta = 0.9999$. We also fix the path loss exponent as $\eta = 3$, which usually ranges between $2.7-3.5$ for urban outdoor scenarios and between $1.6-3.3$ for indoor scenarios~\cite{Goldsmith2005}. Moreover, from Friis equation, we assume $\mathrm{PL}_{\mathrm{dB}}(R_0) = 20 \log_{10} \frac{\lambda}{4\pi R_0}$ and for the simulation we fix $R_0 = 1$ meter and $\lambda = 0.125$ meters, from the industrial, scientific and medical (ISM) radio bands at 2.4 GHz.

\begin{figure}[!t]
\centering
\includegraphics[width= \columnwidth]{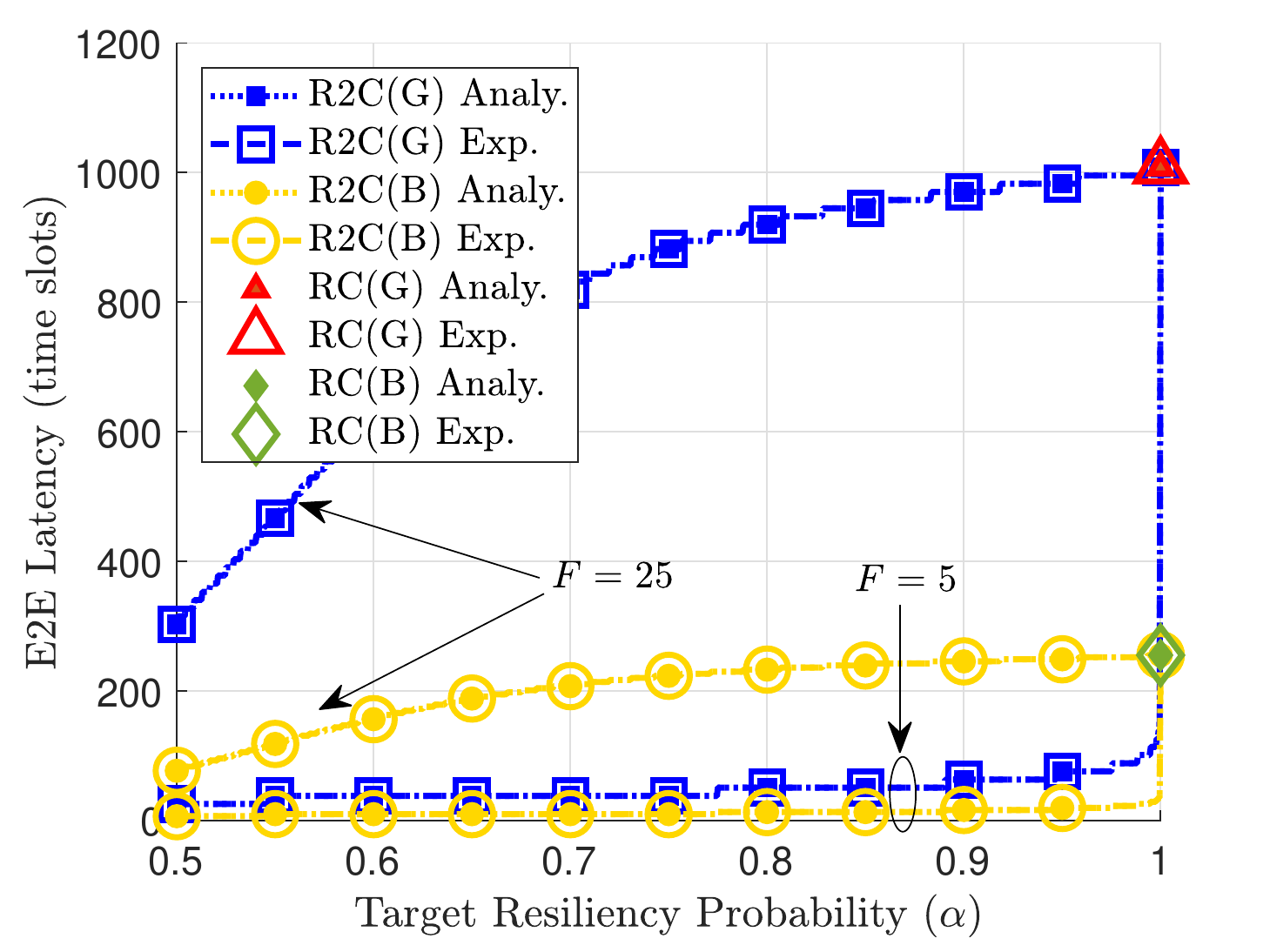}
\caption{\small  The E2E latency of the RC and R2C with gossip (G) and broadcast (B) transmissions versus target resiliency probability $\alpha$, when $F = 5, 25$.}
\label{fig:latencyvsalpha}
\end{figure}

Fig. \ref{fig:latencyvsalpha} illustrates the E2E latency of the RC and R2C with the gossip (denoted by (G)) and broadcast (denoted by (B)) transmissions with respect to the target resiliency probability $\alpha$. From the figure, we can easily see the trade-off between the latency and the resiliency of the R2C. For achieving $\alpha$ close to 1, the number of representative validators $\tilde{N}$ in the R2C must be as large as $N$. Another interesting feature is that when utilizing broadcast transmission, the increment of E2E latency is smaller than that of the gossip protocol. This reveals the broadcast transmission enables low latency consensus while guaranteeing a small loss of resiliency against faulty nodes.

\begin{figure*}[!t]
\centering
\begin{subfigure}{0.495\textwidth}
  \centering
  \includegraphics[width=\textwidth]{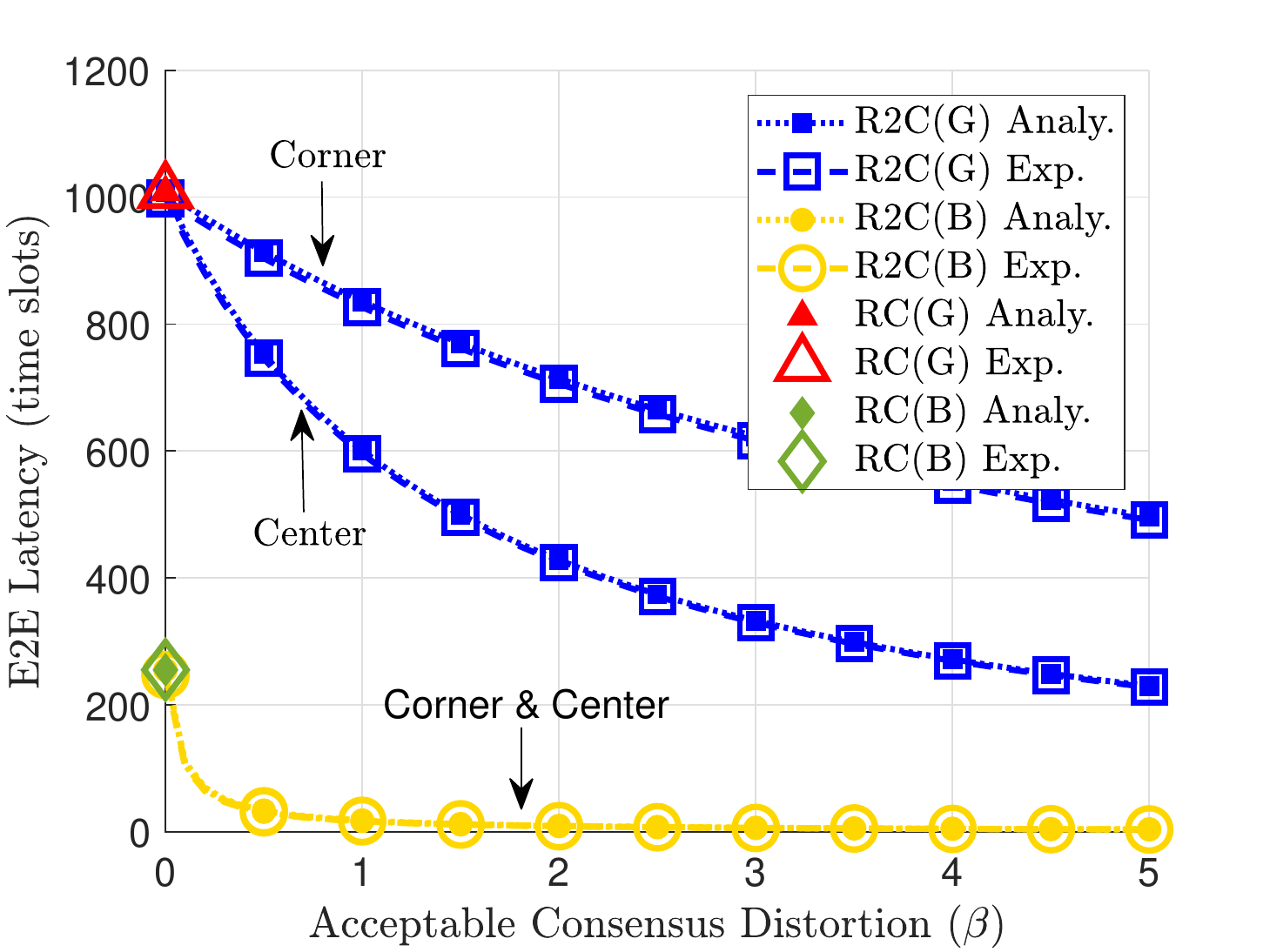}
  \caption{}
  \label{fig:sub1}
\end{subfigure}
\begin{subfigure}{0.495\textwidth}
  \centering
  \includegraphics[width=\textwidth]{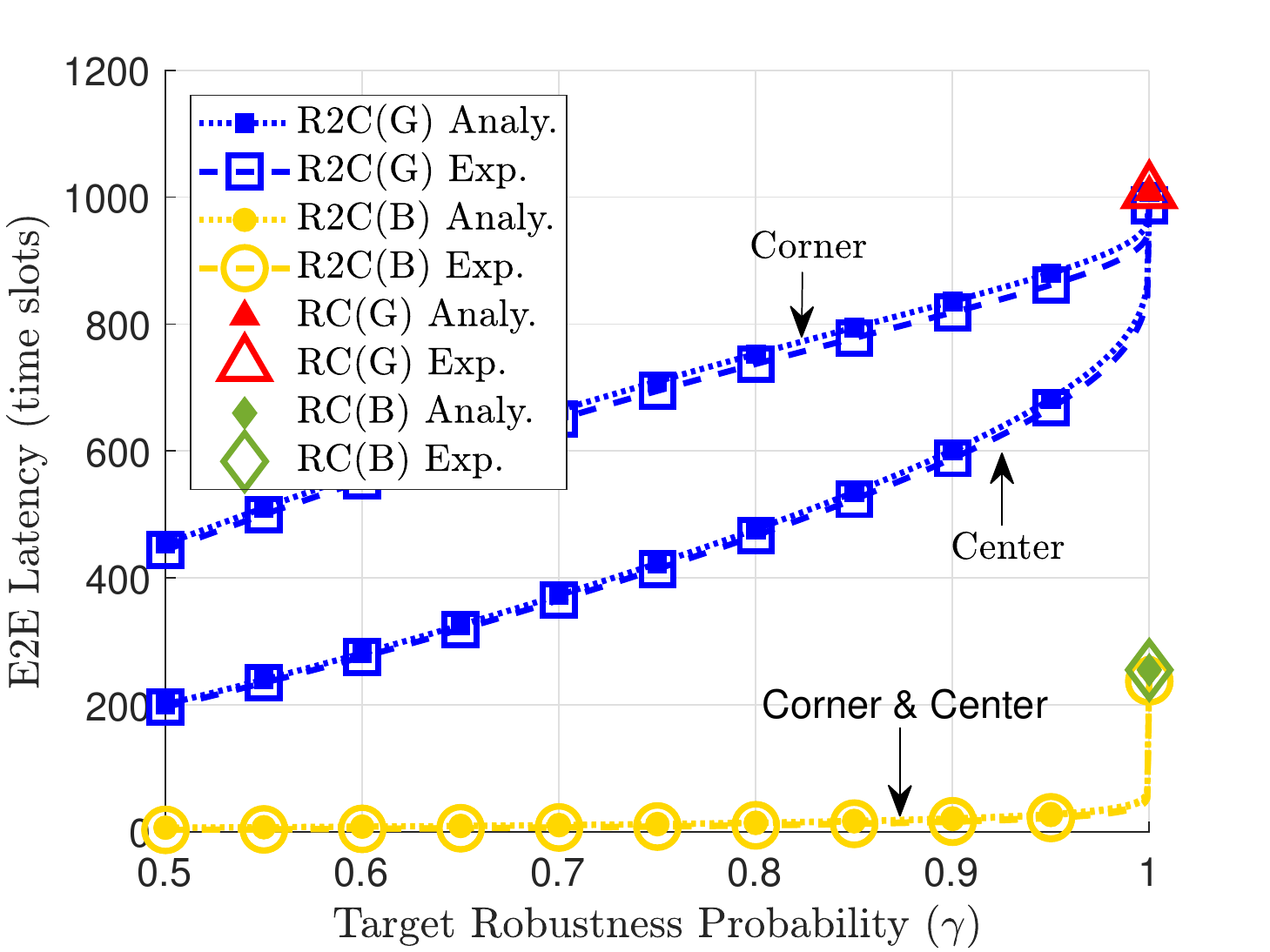}
  \caption{}
  \label{fig:sub2}
\end{subfigure}
\caption{\small  The E2E latency of the R2C with gossip and broadcast transmissions versus (a) the target robustness probability ($\gamma$) with fixed acceptable consensus distortion ($\beta = 1$), and (b) the target robustness probability ($\gamma$) with fixed acceptable consensus distortion ($\beta = 1$).}
\label{fig:latencyenergya}
\end{figure*}

Fig. \ref{fig:sub1} and \ref{fig:sub2} illustrate the E2E latency of the RC and R2C with the gossip and broadcast transmissions with respect to the acceptable consensus distortion and target robustness probability, respectively, when the proposer is located at the corner and center, with $N+1 = 81$ and the target SNR $\rho = 10$ dB. In Fig. \ref{fig:sub1}, we can see that as the acceptable consensus distortion gets smaller, the R2C incur larger delay since it requires a larger number of representative validators. The R2C jointly designed with broadcast transmission outperforms other designs in terms of achieving low E2E latency with small acceptable distortion. Similarly, in Fig. \ref{fig:sub2}, co-design of the R2C with broadcast  transmission can achieve the lowest E2E latency, while guaranteeing robustness more than any other approaches.

\begin{figure*}[!t]
\centering
\begin{subfigure}{0.495\textwidth}
  \centering
  \includegraphics[width=\textwidth]{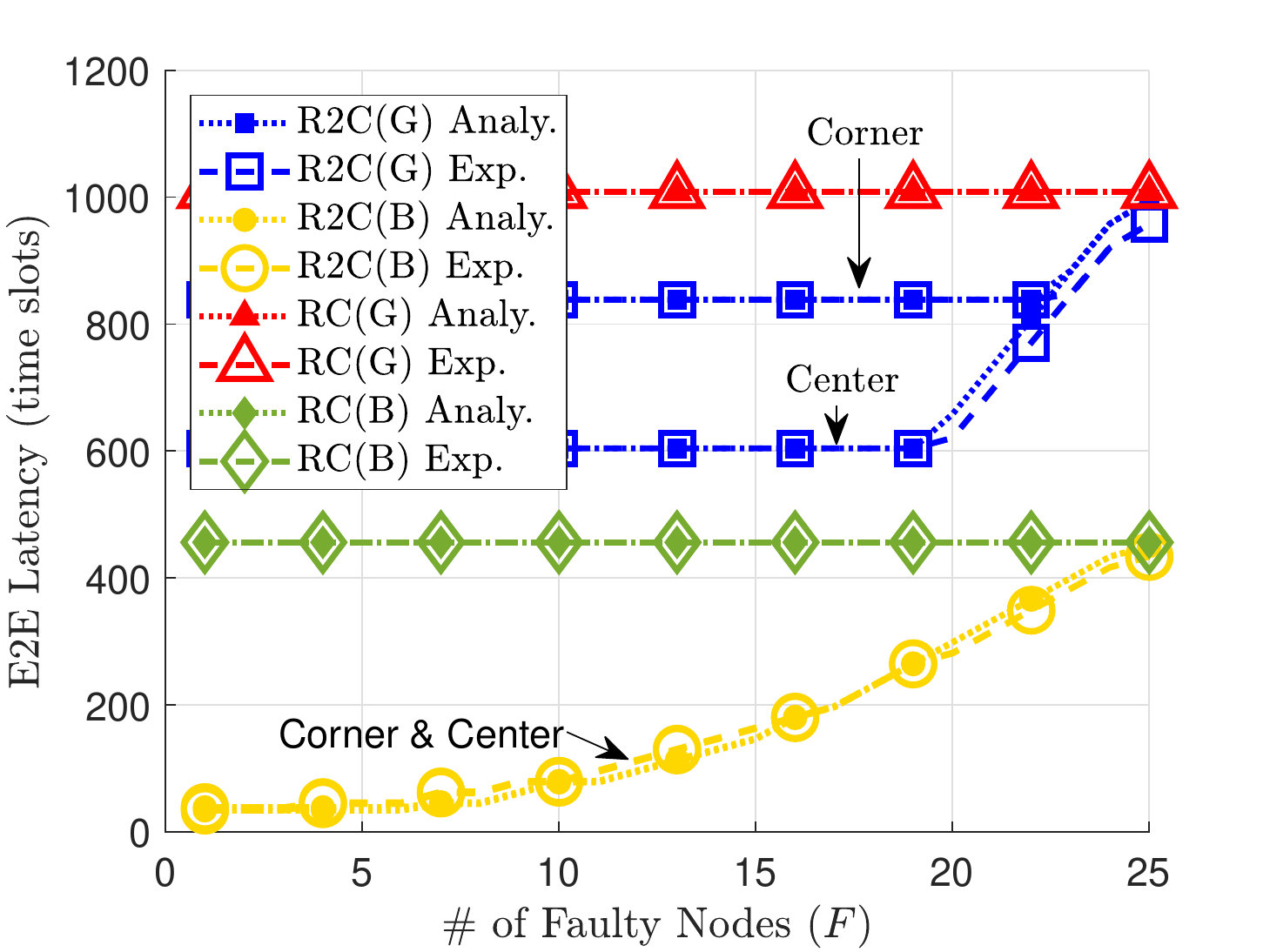}
  \caption{}
  \label{fig:sub3}
\end{subfigure}
\begin{subfigure}{0.495\textwidth}
  \centering
  \includegraphics[width=\textwidth]{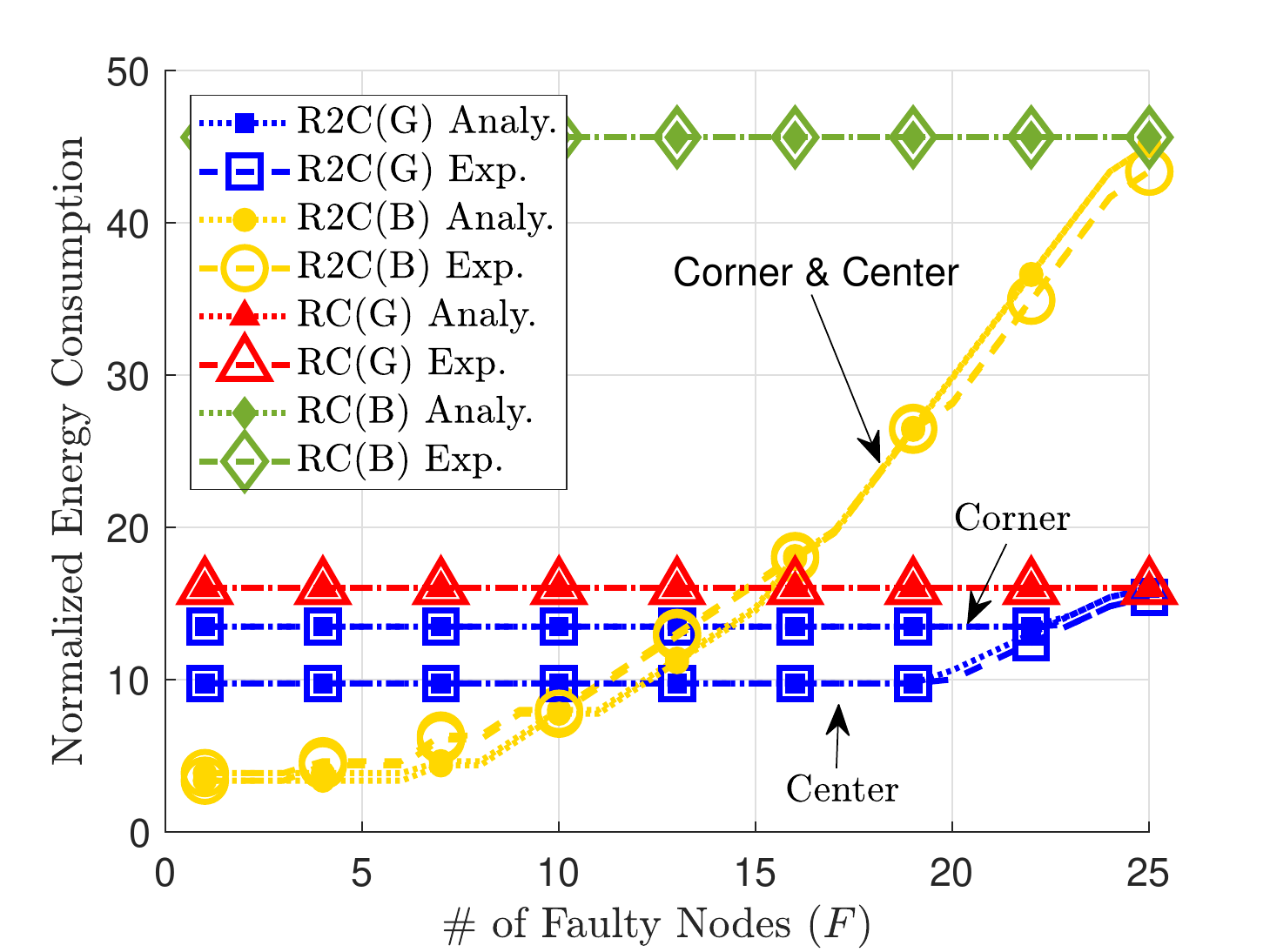}
  \caption{}
  \label{fig:sub4}
\end{subfigure}
\caption{\small (a) The E2E latency and (b) normalized energy consumption versus the number of faulty nodes $F$ of the RC, and $0.99$-resilience and $(1,0.9)$-robustness R2C with gossip and broadcast transmissions.}
\label{fig:latencyenergya}
\end{figure*}

Fig. \ref{fig:sub3}  and \ref{fig:sub4} illustrate the E2E latency and corresponding normalized energy consumption versus the number of faulty nodes $F$, respectively, of the RC and R2C with the gossip and broadcast transmissions when the proposer is located at the corner and the center point of the network, respectively. The reliability factors are fixed to $\alpha = 0.01$, $\beta = 1$, and $\gamma = 0.1$. As shown in Fig. \ref{fig:sub3}, the E2E latency of the R2C is relatively lower than that of the RC, while the R2C with broadcast-based message dissemination can further reduce latency better than using the gossip approach. Note that for small number of faulty nodes, the dominant factor that determines the E2E latency of the R2C becomes guaranteeing $(\beta,\gamma)$-robustness, while for large number of faulty nodes, guaranteeing $\alpha$-resiliency is the dominant factor. In the mean time, Fig. \ref{fig:sub4} shows that the total energy consumption is much larger when using broadcast approach in the RC, however, if there are small number of faulty nodes in the network, the R2C with broadcast transmission can dramatically reduce energy consumption.

\begin{figure}[!t]
\centering
\includegraphics[width=.495\textwidth]{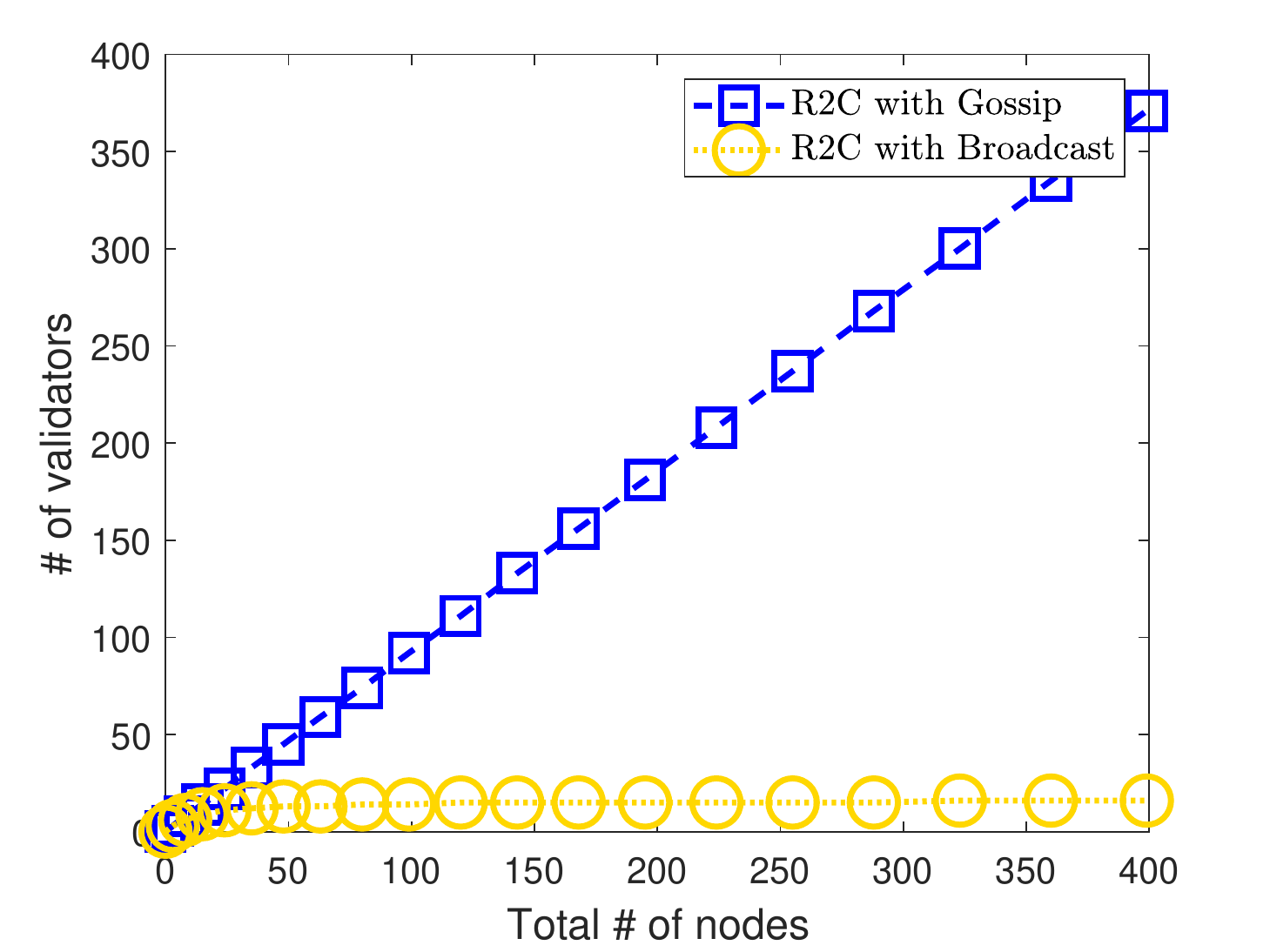}\\
\caption{\small The total number of nodes in the network versus the number of required validators for achieving $\alpha = 0.99$, $\beta = 1$, $\gamma = 0.9$ in the gossip-based and broadcast-based R2C.}
\label{Fig_scalability}
\end{figure}

Finally, Fig. \ref{Fig_scalability} illustrates the total number of nodes in the network versus the number of required validators achieving $\alpha = 0.99$, $\beta = 1$, $\gamma = 0.9$ in the gossip-based and broadcast-based R2C. The network size is fixed to 10,000 square-meters for ensuring direct communications between two nodes located farthest apart. One out of ten nodes is assumed to be a faulty node, so the number of faulty nodes in the network is ten percent. As shown in the figure, the number of validators required in the gossip-based R2C linearly grows as the total number of nodes grows. However, the number of required validators converges to a constant number as the total number of nodes grows. This means that the broadcast-based R2C is scalable and can support as many nodes as possible while using a small number of validators.

\section{Conclusion}
Towards supporting mission-critical and real-time controls in distributed systems, we proposed a novel communication-efficient distributed consensus protocol, i.e., Random Representative Consensus (R2C). For both the gossip and broadcast transmissions, we derived the closed-form expressions of the E2E latency and reliability of the R2C. These expressions bear fundamental relationships between consensus latency and reliability under wireless connectivity, thereby providing a guideline on co-designing distributed consensus and wireless communication protocols. The effectiveness of the R2C was validated numerically and theoretically, for two different network topologies assuming uniformly distributed faulty nodes.

\appendices
\section{Proof of Proposition \ref{prop:e2egossipRC}}\label{app:proofofe2egossipRC}
Consider a message delivery from an arbitrary source node $i$ to destination node $k$, where $i,k \in \mathcal{N}$, $i \neq k$, and $k$ is not necessarily be a neighbor of node $i$. In graph theory, a \emph{walk} is a finite or infinite sequence of edges which join a sequence of nodes and a path is a walk in which all nodes and vertices are distinct. In this perspective, a route of the message delivery from node $i$ to $k$ can be seen as a path between the two nodes. A single message may flow over various paths, while we focus on the one with the shortest amount of time upon all successful message delivery. Let $Z_{g,ik}$ be a random variable which denotes the least number of time slots required to deliver a message from source $i$ to destination $k$ via the gossip method. Among all the possible paths for the message flow, we define the shortest paths as the paths which are comprised of the minimum number of edges. We also define the number of edges comprising the shortest paths between the node $i$ and $k$ as $e_{ik}$ and the number of shortest paths between the two nodes as $s_{ik}$. Intuitively, the shortest paths are the dominant factors that determines $Z_{g,ik}$, if there are sufficiently large number of independent shortest paths from node $i$ to node $k$. Furthermore, a convergence of $Z_{g,ik}$ to $e_{ik}$ if $s_{ik} \rightarrow \infty$ is intuitive trivial, since at least one shortest path without any outage will exist among $s_{ik}$ shortest paths.
\begin{mylemma}\label{lem:pmf_gossip_T}
Given a source-destination node pair $(i,k)$, the random variable $Z_{g,ik}$ converges in distribution to a constant random variable as 
  \begin{align}\label{eq:pdf_T}
\Pr\left[Z_{g,ik} = z\right] = \begin{cases} 1, & z = e_{ik}\\ 0, & \text{elsewhere.}\end{cases}
\end{align} 
if the shortest paths do not have any internal edge in common, i.e., edge-independent, and $s_{ik}\rightarrow \infty$.
\end{mylemma}
\begin{IEEEproof}
The outage probability of a single hop communication between two neighbors can be written as
  \begin{align}\label{eq:snroutagegossip}
\epsilon_{g} = 1- \exp\left( - 10^{\frac{\mathrm{PL}_{\mathrm{dB}}(R_0)}{10}} \rho\frac{P_{n}}{P_{g,t}} \left(\frac{R}{R_0}\right)^{\eta} \right),
\end{align} 
from \eqref{eq:snroutage}. Suppose there exist $s_{ik}$ shortest paths between node $i$ and $k$ and let $T_{1}, \dots, T_{s_{ik}}$ be the random variables that denote the number of transmissions required for a successful message delivery over those shortest paths. Then, we can model $T_{s}$ as a random variable which follows the negative binomial distribution as
  \begin{align} \label{eq:singlepath}
\Pr\left[T_{s} = t \right] = \begin{cases} \binom{t  - 1}{e_{ik} - 1} \epsilon_g^{t - e_{ik}} (1 - \epsilon_g)^{e_{ik}}, & t \geq e_{ik},\\
0, & t < e_{ik}.\end{cases}
\end{align} 
for $s \in \{1,\dots,s_{ik}\}$, where $t$ is a non-negative integer.

Let $U_{ik}$ be a random variable which denotes the minimum number of transmissions for successful message delivery from node $i$ to node $k$, over all the paths from node $i$ to node $k$ excluding the shortest paths. We assume that $U_{ik}$ is following some p.m.f. $\Pr[U_{ik} = u]$, which is obviously $\Pr[U_{ik} = u] = 0$ for $u \leq e_{ik}$, since no paths without the shortest paths can deliver a message with less than or equal to $e_{ik}$ transmissions. Then we can write 
  \begin{align}
Z_{g,ik} = \min \{T_{1}, \dots, T_{s_{ik}}, U_{ik}\},
\end{align} 
where the random variables $T_{1}, T_{2}, \dots, T_{s_{ik}}$ are identically and independently distributed following p.m.f. of the negative binomial distribution \eqref{eq:singlepath} with parameter $e_{ik}$. Then the cumulative distribution function (c.d.f.) of $Z_{g,ik}$ can be derived as
  \begin{align}
\Pr\left[Z_{g,ik} \! \leq \! z \right] \! &= \! 1 - \Pr \left[Z_{g,ik}> z\right]\\
&= \! 1 - \Pr \left[ \min \{T_{1}, \dots, T_{s_{ik}},U_{ik}\} >z \right]\\
&=\! 1 - \Pr \left[ T_{1} > z, \dots, T_{s_{ik}} > z,U_{ik}>z \right]\\
\label{eq:independency}&= \! 1 \! - \! \big(1\!-\!\Pr[U_{ik} \leq z] \big)\left( 1 \! - \! \Pr[T_s \leq z] \right)^{s_{ik}},
\end{align} 
where \eqref{eq:independency} follows from the assumption that all the paths are independent.
If there exist a sufficiently large number of shortest paths, i.e., $s_{ik} \rightarrow \infty$, we have
  \begin{align}\label{eq:limitCDF_T}
\lim_{s_{ik} \rightarrow \infty} \Pr[Z_{g,ik} \leq z] = \begin{cases} 0, & z < e_{ik},\\
1, & z \geq e_{ik},
\end{cases}
\end{align} 
since $\Pr[U_{ik} \leq z] = 0$ and $\Pr[T_{s} \leq z] = 0$ for $z < e_{ik}$, and $0 <\Pr[T_{s} \leq z]< 1$ for $z \leq e_{ik}$.
From \eqref{eq:independency}, we also have
  \begin{align}
\Pr[Z_{g,ik} = z] &= \left(\Pr[Z_{g,ik} \leq z] - \Pr[Z_{g,ik} \leq z-1]\right).
\end{align} 
Thus, from \eqref{eq:limitCDF_T} we have \eqref{eq:pdf_T} and this completes the proof of Lemma \ref{lem:pmf_gossip_T}.
\end{IEEEproof}
Put differently, if there are infinitely many edge-independent shortest paths from the node $i$ to $k$, there may exist at least one path that can guarantee a successful message delivery from the node $i$ to $k$ without occurring any outages during the delivery. In fact, the total number of shortest paths $s_{ik}$ is finite and the number of shortest paths that are edge-independent is much smaller than $s_{ik}$. Therefore, the random variable $Z_{g,ik}$ can be lower bounded as
  \begin{align}\label{eq:lowerboundZgik}
Z_{g,ik} \geq e_{ik},
\end{align} 
in the considered network model. From \eqref{eq:lowerboundZgik}, we can also conclude that 
  \begin{align}\label{eq:lowerboundmaxZgik}
w_{i,\zeta} &\geq \max_{k} e_{ik},
\end{align} 
for all $i \in \mathcal{N}$.
For small $\epsilon_g$, the bound \eqref{eq:lowerboundmaxZgik} tight.
For sufficiently small $\epsilon_g$, we approximately get
\begin{align}
\Pr\left[ \max_{k} Z_{g,ik} > \max_{k} e_{ik} \right] \approx 0.
\end{align} 

In the mean time, in the square network considered in this paper, the number of edges that comprises shortest paths is $e_{ik} = \tilde{x}_{ik} + \tilde{y}_{ik}$, where $\tilde{x}_{ik} = |x_{k} - x_{i}|/R$ and $\tilde{y}_{ik} = |x_{k} - y_{i}|/R$, and from simple combinatorics the number of shortest paths between the two nodes can be readily derived as $s_{ik} = \frac{\left( \tilde{x}_{ik} + \tilde{y}_{ik}\right)!}{\tilde{x}_{ik}! \tilde{y}_{ik}!}$. Since we assume square network composed of $N+1$ nodes, and from \eqref{eq:lowerboundmaxZgik} and \eqref{eq:baselinelatency}, we have \eqref{eq:Lg}.


\section{Proof of Proposition \ref{prop:e2ebroadcastRC}}\label{app:proofe2ebroadcastRC}
Let $Z_{b,ik}$ be a random variable which denotes the number of time slots required to deliver a message from source node $i$ to destination node $k$ via the broadcast protocol. In this case, messages are delivered in a single hop without any help from relays. Since the messages are sent repeatedly until successful delivery, the random variable $Z_{b,ik}$ follows the geometric distribution with p.d.f.
\begin{align} \label{eq:pmf_Tbik}
\Pr[Z_{b,ik} = z] = \begin{cases}\epsilon_{ik}^{z-1} \left(1 - \epsilon_{ik} \right), & \forall z \geq 1\\
0, & \mathrm{elsewhere},
\end{cases}
\end{align}
where $\epsilon_{ik} = \Pr\left[ \mathrm{SNR}_{ik} < \rho \right]$ is the SNR outage probability \eqref{eq:snroutage} with transmit power $P_t = P_{t,b}$. Then the dissemination outage probability can be upper bounded as
\begin{align}
\Pr\left[\max_{k}Z_{b,ik} > w_{i,\zeta}\right] &= 1 - \Pr\left[\max_{k}Z_{b,ik} \leq w_{i}\right]\\
\label{eq:independentZbik}&= 1 - \prod_{ k \in \mathcal{N}, k \neq i } \Pr\left[Z_{b,ik} \leq w_{i} \right]\\
&= 1 -  \prod_{ k \in \mathcal{N}, k \neq i } \left( 1 - \epsilon_{ik}^{w_{i}} \right)\\
\label{eq:lowerboundTbik}&\leq 1 - \left( 1 - \epsilon_{i,\max}^{w_{i}} \right)^{N},
\end{align}
where $\epsilon_{i,\max} = 1 - \exp\left( - 10^{\frac{\mathrm{PL}_{\mathrm{dB}}(R_0)}{10}} \rho\frac{P_{n}}{P_t} \left(\frac{\max_{k} R_{ik}}{R_0}\right)^{\eta} \right)$ is the SNR outage probability between node $i$ and the node located maximum distance apart from node $i$. The equality \eqref{eq:independentZbik} holds from the independence of the channels between the source and the destinations, and the inequality \eqref{eq:lowerboundTbik} holds from the fact that the SNR outage probability is the largest when the node $k$ is located farthest apart from the node $i$ among all the destinations from \eqref{eq:snroutage}. Since the dissemination time duration achieves a dissemination outage probability smaller than $\zeta$, we have a bound as follows.
\begin{myremark}\label{remark:broadcastwi}
In the broadcast method, the dissemination time duration is bounded as
\begin{align}\label{eq:wbi}
w_{i} \geq \left\lceil \frac{\log\left( 1 \! - \! \zeta^{\frac{1}{N}}\right)}{\log{\epsilon_{i,\max}}} \right\rceil,
\end{align}
since \eqref{eq:lowerboundTbik} must be smaller than or equal to $\zeta$ from \eqref{eq:zeta}.
\end{myremark}
From Remark \ref{remark:broadcastwi} and \eqref{eq:baselinelatency}, we get \eqref{prop:e2ebroadcastRC}.

\section{Proof of Proposition \ref{thm:reliability}}\label{app:proofofbetagamma}
Throughout the proof, we fix a proposer as node $p$ without loss of generality and use $D$ and $T_i$ instead of $D(A_p,\mathcal{V}_p,\tilde{\mathcal{V}}_p)$ and $T(A_p)_i$ from \eqref{eq:distortionandtime}, respectively, for simple notation.

We first show that $\mathrm{E}[D] = 0$. Let $\tilde{\mathcal{V}}_{p} \in \left\{\tilde{\mathcal{V}}_{p,1},\dots,\tilde{\mathcal{V}}_{p,\binom{N}{\tilde{N}}}\right\}$ be the randomly chosen validator set by the proposer, where $\tilde{\mathcal{V}}_{p,1},\dots, \tilde{\mathcal{V}}_{p,\binom{N}{\tilde{N}}}$ be all possible sets of cardnality $\tilde{N}$ that can be chosen from $\mathcal{V}_p$ with equal probability ${1}/{\binom{N}{\tilde{N}}}$. For given $\tilde{\mathcal{V}}_p = \tilde{\mathcal{V}}_{p,l}$, we have
\begin{align}
\mathrm{E}\left[D \! \mid \! \tilde{\mathcal{V}}_{p,l}\right] &= \mathrm{E}\left[\frac{1}{N} \sum_{i \in \mathcal{V}_p} T_{i} - \frac{1}{\tilde{N}} \sum_{j \in \tilde{\mathcal{V}}_{p,l}} T_{j}\right]\\
&= \mathrm{E}\left[\frac{1}{N} \sum_{i \in \tilde{\mathcal{V}}_{p,l}^{c}} \! T_{i} \! - \! \frac{N \! - \! \tilde{N}}{\tilde{N}N} \sum_{j \in \tilde{\mathcal{V}}_{p,l}} \! T_{j}\right]\\
\label{eq:Cdifferenceconditional} &= \frac{1}{N} \sum_{i \in \tilde{\mathcal{V}}_{i,l}^{c}} \mathrm{E}[T_{i}] - \frac{N- \tilde{N}}{\tilde{N}N} \sum_{j \in \tilde{\mathcal{V}}_{i,l}} \mathrm{E}[T_{j}],
\end{align}
where $\tilde{\mathcal{V}}_{i,l}^{c} =\mathcal{V}_{i} \backslash \tilde{\mathcal{V}}_{i,l}$ is the complement of $\tilde{\mathcal{V}}_{i,l}$. By marginalizing \eqref{eq:Cdifferenceconditional}, we have
\begin{align}
\mathrm{E}[D] \! &= \! \frac{1}{\binom{N}{\tilde{N}}} \sum_{l = 1}^{\binom{N}{\tilde{N}}} \mathrm{E}\left[D \mid \tilde{\mathcal{V}}_{p,l}\right]\\
\! &= \! \frac{1}{\binom{N}{\tilde{N}}}\sum_{l = 1}^{\binom{N}{\tilde{N}}}\left(  \frac{1}{N} \! \sum_{i \in \tilde{\mathcal{V}}_{p,l}^{c}} \! \mathrm{E}[T_{i}] \! - \! \frac{N \! - \! \tilde{N}}{\tilde{N}N} \! \sum_{j \in \tilde{\mathcal{V}}_{p,l}} \! \mathrm{E}[T_{j}] \right)\\
&= \frac{N\!-\! \tilde{N}}{\binom{N}{\tilde{N}}} \left(  \frac{1}{N} \sum_{i \in \mathcal{V}_{p}} \mathrm{E}[T_{i}] - \frac{1}{\tilde{N}N} \sum_{j \in \mathcal{V}_{p}} \tilde{N} \mathrm{E}[T_{j}] \right)\\
&= 0.
\end{align}

From Chebyshev's inequality, we have
\begin{align}
\Pr\left[ | D | \geq \sqrt{\beta} \right] \leq \frac{\mathrm{Var}(D)}{\beta}. \label{eq:chebyvaru},
\end{align}
where from the law of total variance
\begin{align}\label{eq:lawoftotalvariance}
\mathrm{Var}(D) &= \mathrm{E}\left[\mathrm{Var}\left(D \mid \tilde{\mathcal{V}}_{p}\right)\right] +  \mathrm{Var}\left(\mathrm{E}\left[D \mid \tilde{\mathcal{V}}_{p}\right]\right).
\end{align}

The first term of the right-hand side in \eqref{eq:lawoftotalvariance} can be rewritten as
\begin{align}
\nonumber&\mathrm{E}\left[\mathrm{Var}\left(D \! \mid \! \tilde{\mathcal{V}}_{i}\right)\right] \\
&= \! \frac{1}{\binom{N}{\tilde{N}}}\sum_{l = 1}^{\binom{N}{\tilde{N}}} \mathrm{Var} \left( \frac{1}{N} \sum_{i \in \tilde{\mathcal{V}}_{p,l}^{c}} T_{i} - \frac{N- \tilde{N}}{\tilde{N}N} \sum_{j \in \tilde{\mathcal{V}}_{p,l}} T_{j} \right)\\
&= \frac{1}{N^2\binom{N}{\tilde{N}}}\sum_{l = 1}^{\binom{N}{\tilde{N}}} \left( \sum_{i \in \tilde{\mathcal{V}}_{p,l}^{c}} \! \mathrm{Var}(T_{i}) \! + \! \frac{(N\!-\!\tilde{N})^2}{\tilde{N}^2} \! \sum_{j \in \tilde{\mathcal{V}}_{p,l}} \! \mathrm{Var}(T_{j}) \right)\\
&= \frac{1}{N^2} \sum_{i \in \mathcal{V}_p} \left( \frac{(N\!-\!\tilde{N})}{N} \mathrm{Var}(T_{i}) + \frac{\tilde{N}(N-\tilde{N})^2}{\tilde{N}^2 N^3} \mathrm{Var}(T_{i})\right)\\
&= \frac{(N-\tilde{N})}{\tilde{N}N^2} \sum_{i \in \mathcal{V}_p}\mathrm{Var}(T_{i}), \label{eq:EVarD}
\end{align}
and the second term as
\begin{align}
\nonumber&\mathrm{Var}\left(\mathrm{E}\left[D\mid \tilde{\mathcal{V}}_i\right]\right) \\
&=\! \frac{1}{\binom{N}{\tilde{N}}} \sum_{l = 1}^{\binom{N}{\tilde{N}}} \left( \frac{1}{N} \sum_{i \in \tilde{\mathcal{V}}_{p,l}^{c}} \mathrm{E}[T_{i}] \! - \! \frac{N\!-\!\tilde{N}}{\tilde{N}N} \sum_{j \in \tilde{\mathcal{V}}_{p,l}} \mathrm{E}[T_{j}] \right)^2\\
&=\! \frac{(N-\tilde{N})}{\tilde{N}N^2}\sum_{i \in \mathcal{V}_p} \left( \mathrm{E}[T_{i}]^2 \! - \! \frac{1}{(N\!-\!1)}\sum_{j \in \mathcal{V}_p, j \neq i} \mathrm{E}[T_{i}]\mathrm{E}[T_{j}]\right). \label{eq:VarED}
\end{align}

From \eqref{eq:lawoftotalvariance}, \eqref{eq:EVarD} and \eqref{eq:VarED}, and since $\mathrm{Var}(T_{i}) = \mathrm{E}[T_{i}^2] - \mathrm{E}[T_{i}]^2$, we have
\begin{align}\label{eq:varu}
\nonumber&\mathrm{Var}(D)\\
& = \frac{\tau^2(N\!-\!\tilde{N})}{\tilde{N}N^2} \sum_{i\in \mathcal{V}_p} \left( \mathrm{E}[T_{i}^2] \! - \! \frac{1}{N-1} \sum_{j \in \mathcal{V}_p, j \neq i} \mathrm{E}[Z_{i}] \mathrm{E}[Z_{j}]\right).
\end{align}
By substituting \eqref{eq:varu} into \eqref{eq:chebyvaru}, we have \eqref{eq:D}. This finishes the proof.


\bibliographystyle{ieeetr}  
\bibliography{IEEEabrv,Blockchain}

\end{document}